\documentclass[11pt]{article}

\usepackage{amsmath,amssymb}
\usepackage[normalem]{ulem}
\usepackage{booktabs}
\usepackage{sidecap}
\usepackage{comment}
\usepackage[labelformat=simple]{subcaption}

\DeclareCaptionLabelFormat{subcaptionlabel}{\normalfont(\textbf{#2}\normalfont)}
\usepackage{changepage}
\usepackage{array}
\usepackage{longtable}
\usepackage{geometry}
\usepackage{graphicx}
\usepackage{tikz}
\usetikzlibrary{calc,shapes.geometric,arrows.meta,positioning}
\usepackage{algorithm}
\usepackage{algpseudocode}
\usepackage[hidelinks]{hyperref}

\usepackage{xcolor}
\newcommand{\orcid}[1]{%
  \href{https://orcid.org/#1}{\textcolor{green}{\small\texttt{ORCID: #1}}}%
}

\title{Shared Nodes of Overlapping Communities in Complex Networks}

\author{%
Vesa Kuikka\ \orcid{0000-0002-3677-816X}$^{*}$, 
Kosti Koistinen\ \orcid{0009-0007-9459-6309}, \\
Kimmo K. Kaski\ \orcid{0000-0002-3805-9687} \\[0.5em]
Department of Computer Science, Aalto University School of Science\\
P.O. Box 15500, 00076 Aalto, Finland
}

\date{}

\begin{document}

\maketitle

\begin{abstract}
Overlapping communities are key characteristics of the structure and function analysis of complex networks. Shared or overlapping nodes within overlapping communities can form either subcommunities or act as intersections between larger communities. Nodes at the intersections that do not form subcommunities can be identified as overlapping nodes or as part of an internal structure of nested communities. To identify overlapping nodes, we apply a threshold rule based on the number of nodes in the nested structure. As the threshold value increases, the number of selected overlapping nodes decreases. This approach allows us to analyse the roles of nodes considered overlapping according to selection criteria, for example to reduce the effect of noise. We illustrate our method by using three small and two larger real-world network structures. In larger networks, minor disturbances can produce a multitude of slightly different solutions, but the core communities remain robust, allowing other variations to be treated as noise. While this study employs our own method for community detection, other approaches can also be applied. Exploring the properties of shared nodes in overlapping communities of complex networks is a novel area of research with diverse applications in social network analysis, cybersecurity, and other fields in network science.
\end{abstract}


\section{Introduction}
\label{sec:Introduction}

Detecting communities in complex networks is an essential aspect of analysing network structures, as it helps identify clusters or modules of complex systems, such as social and biological networks, and cyber-physical systems. In networks, a community consists of nodes that are tightly linked to each other because of shared attributes. In social contexts, members of a community typically interact with each other more frequently than with individuals outside their group. Several reviews on community detection methodologies have been published in the literature~\cite{Souravlas,LI}. In this study, our objective is not to analyse or compare different community detection methods but instead to treat the detected communities and their structures as given.

Communities can be categorised as non-overlapping, overlapping, or hierarchical. Non-overlapping communities assign each node to exactly one community, reflecting structures where the roles or associations of nodes are mutually exclusive. In contrast, overlapping communities allow nodes to belong simultaneously to multiple communities, as in social networks as members of different social circles, such as family, school, hobbies, etc~\cite{barabasi, Newman, Latapy2008}. Hierarchical communities present a nested structure, with communities organised at multiple levels such that larger groups encompass smaller subcommunities, revealing multi-scale patterns within the network. These distinctions are crucial for tailoring community detection methods to the specific characteristics of real-world networks~\cite{Newman}.

In network communities, the overlapping nodes play a crucial role in the structure and function of the network, as they frequently act as bridges that facilitate the flow of information or influence between different communities~\cite{Shang2015}. Identifying and analysing overlapping nodes is therefore essential to understand how complex systems operate. Studying them can reveal, e.g., critical points of vulnerability in infrastructure networks, or improve predictions of spreading processes, such as opinion diffusion or contagion \cite{Ghalmane2019,Yang2012}. Moreover, overlapping nodes tend to exert stronger spreading influence than their non-overlapping counterparts. \cite{Koistinen2025OverlappingNodes, diffusionhur2012}. In summary, overlapping nodes reveal how local communities connect, cooperate, and collectively shape the behaviour of the network as a whole \cite{Ghalmane2019}.

Overlapping community detection methods can identify nodes that participate in multiple groups, but they face several limitations in handling such overlaps. It is often difficult to distinguish genuine multi-membership from artefacts caused by noise or high-degree nodes, and there is no universally accepted definition of what constitutes meaningful overlap. The resulting assignments can be sensitive to parameter choices and heavy overlap can reduce the interpretability of the detected structure. Moreover, validating overlapping communities remains challenging due to the lack of clear ground truth, thus complicating the assessment of the reliability of overlapping node assignments. These limitations are well documented in reviews of community detection methods, highlighting issues such as ambiguous definitions of overlap, sensitivity to parameters, challenges in validation, and difficulties in separating true overlaps from structural noise~\cite{Fortunato2010,xie2013overlapping,yang2016comparative,Fortunato}.

In order to identify the building blocks that form network communities, we apply the method presented in our previous study~\cite{KuikkaCD}. Here, we will demonstrate how small and cohesive substructures can be used to identify shared or overlapping nodes belonging to multiple communities.  At the intersections of these communities, there are shared nodes, which, according to our definition, do not form distinct subcommunities. By applying a sequential selection rule to these nodes, we can distinguish the remaining overlapping nodes from those that make up the internal structure of nested communities. We also discuss the potential effects of ambiguously assigned nodes, i.e., noise, that can occur in detected communities. By applying threshold rules, we can determine which nodes to retain as overlapping nodes and which ones to exclude.

In complex networks, inherently overlapping structures and dense interconnections between communities can be found, making it necessary to apply filtering to detect and isolate the most influential overlapping nodes~\cite{Shang2015}. However, detecting overlapping nodes and distinguishing them from noise has received limited attention, in part because there is no universally accepted definition of what constitutes a community~\cite{FortunatoHric2016}. Here, our aim is to improve the reliability of community structure analysis and provide a deeper understanding of how nodes interact and influence each other. Regarding the community detection method, our approach to analyse shared and overlapping nodes can also be applied to other community detection methods, provided that they can identify overlapping communities.

In the present study, we compare the number of nodes in the outer layer with the inner layer of the nested structures. The overlapping nodes are selected by comparing this ratio with a threshold parameter value. As the threshold value increases, the criterion becomes stricter, resulting in more exclusions and fewer overlapping nodes remaining. In the end, increasing the threshold can lead to only one overlapping node. In the case of larger networks, minor disturbances can result in numerous slightly varied theoretical solutions~\cite{leskovec2009community}. However, core communities typically maintain their stability, indicating the robustness of our method. Our results show that for all network sizes, the number of overlapping nodes gradually decreases as the threshold parameter increases. This illustrates a key property of our model, namely that it can effectively limit or exclude potential overlapping nodes, even when community detection methods initially identify large sets of overlapping communities~\cite{kuikka2024detailed}.

In this study, we apply our method to five real-world networks. For three of them, we compared empirical observations with the theoretical results on overlapping nodes derived from our model. In all cases, the findings indicate a trend that as the threshold value of the selection rule increases, the number of selected overlapping nodes decreases. This suggests that the influence of these overlapping nodes grows both in empirical observations and in theoretical calculations. Additionally, two of the example cases illustrate our model's performance in larger network structures where empirical data is unavailable. In these instances, we focus solely on demonstrating how our model works in these larger networks.

The research objective of the present study is to develop a threshold-based method that identifies overlapping nodes in complex networks by filtering out noise-induced overlaps and internal subcommunity structures This allows the analysis to focus on the remaining nodes whose role and influence as overlapping connectors become increasingly pronounced.

The following sections first review related works on overlapping community detection, after which we present our model and the threshold-based selection rule to identify overlapping nodes. We then demonstrate the method using several real-world network datasets and analyse how  thresholding influences the resulting overlaps. Finally, we discuss the implications of the approach and conclude with a summary of the findings and directions for the future research.

\section{Related Work}
\label{sec:Related_Work}

The study of overlapping communities with nodes belonging to multiple groups has become an important topic of network science. Early community detection methods assumed disjoint community partitions, but subsequent research revealed that social, biological, and information networks frequently exhibit shared or ambiguous nodal memberships. One of the first major contributions was made in~\cite{Palla2005}, which introduced the clique percolation method, allowing communities to overlap through shared nodes. Later, the study in~\cite{Ahn2010} proposed link communities, shifting the focus from nodes to edges to better capture multi-scale relationships. The study in \cite{lancichinetti2011finding} introduced a local statistical fitness optimisation framework to detect hierarchical and overlapping communities. It evaluates candidate communities against a null model (pseudo-community) that preserves the degree distribution of the network, allowing statistically significant structures to be distinguished from noise. Other probabilistic models, such as the mixed membership stochastic block model in~\cite{Airoldi2008}, further formalised overlapping membership of nodes through statistical inference. Complementary structural approaches address the issue of high overlap with merging and noise filtering. For example, in \cite{Xu2019Uncovering}, a model identifies partial community cores, merges those that represent the same underlying community, and subsequently treats the remaining weakly affiliated nodes as noise. However, these approaches tend to struggle with hierarchical or nested structures, which can cause them to overestimate overlap or misinterpret subcommunities as genuine multi-memberships.

More recently, machine learning approaches have been extended to the aforementioned frameworks. Popular methods such as BigCLAM \cite{Yang2013BigCLAM} and CESNA \cite{Yang2012}, and related generative approaches, frame overlapping memberships through learnable statistical parameters. Deep learning has further advanced the field by enabling unsupervised representations of the network structure, for example through autoencoder-based approaches \cite{Hinton2006Reducing_ae, Kingma2014VAE}, as well as generative adversarial approaches specifically designed for overlapping community detection, such as CommunityGAN \cite{CommunityGAN2019}. Subsequently, graph neural network approaches, including Graph Convolutional Networks (GCN) and Graph Attention Networks (GAT), have proven particularly effective, as they propagate information between local neighbourhoods while preserving complex dependency patterns \cite{SISMANIS2025107529}. However, deep learning–based methods often introduce the challenge that their performance being highly dependent on hyperparameters, training heuristics, and model architecture selection. This can make the detection of overlapping nodes less transparent, less reproducible, and more difficult to interpret. Moreover, the learned representations may obscure the structural reasons why particular nodes are classified as overlapping, which makes the models less interpretable. For further details and timeline of machine learning models in community detection, see~\cite{survey2022_cd_ml} for a comprehensive survey. 

Many community detection methods are based on dividing the network into two factions. Another approach is to investigate the structure of the network by dividing it into the core and the periphery~\cite{yanchenko2023}. The core nodes are densely connected to each other, but also have connections to the periphery nodes. In contrast, the periphery nodes have only sparse connections among themselves. An alternative definition could be that the core nodes are located a short distance from all other nodes.

The study presented in~\cite{Yang2012} extends the detection of overlapping communities to include attributed networks by integrating both the characteristics of the nodes and the structure of the network. In~\cite{Rossetti2018Survey}, a comprehensive survey of community detection is provided, covering both overlapping and temporal community structures. Recently, advances have been made in the study of overlapping communities and their intersections. One study~\cite{khawaja2025common} investigates new methods that utilise neighbour similarity metrics to detect overlaps and subtle communities. Another study~\cite{guo2022overlapping} emphasises the importance of tie strength in distinguishing between overlapping and non-overlapping nodes, which is useful for exploring structural criteria related to overlaps. In addition, an adaptive density-based method~\cite{niu2023overlapping} has been introduced to detect overlaps, which employs iterative partitioning and is effective for networks with varying scales. Collectively, these studies highlight the detection of overlapping communities as an essential framework to understand the complex and multifaceted organisation of real-world networks.

Despite these advances, several important research gaps still remain. Existing overlapping community-detection methods identify shared nodes but offer limited guidance on distinguishing true overlapping nodes from noise or internal subcommunity structures, leaving their post-processing largely unresolved \cite{FortunatoHric2016, xie2013overlapping, Shen2009Overlap}. Although structural, probabilistic, and deep-learning approaches provide diverse ways to detect overlaps \cite{Palla2005, Ahn2010, Airoldi2008, Yang2013BigCLAM, survey2022_cd_ml}, the literature still lacks a unified threshold-based framework capable of controlling noise, handling nearly identical community structures, and revealing the increasing influence of the remaining overlapping nodes \cite{riolo, lancichinetti2011finding, CherifiPalla}. Furthermore, evaluation practices remain inconsistent, particularly due to the mismatch between attribute-driven social circles and topological communities, which highlight the need for clearer methodological criteria and more robust analysis of the structural roles of overlapping nodes \cite{Leskovec2012FacebookEgo, Yang2015DefiningCommunities, circles_communities2014}.

\section{Model}
\label{sec:Model}

In Section~\ref{subsec:ISM}, we present an overview of our community detection model~\cite{KuikkaCD}, as it helps to understand the new contributions in the current paper. Our community detection method is effective for demonstrations because it can generate different overlapping structures based on the link weights used in the calculations. Higher link weights improve network cohesion, resulting in the detection of fewer communities and, typically, fewer overlapping communities. In Section~\ref{subsec:ONM}, we focus on identifying and categorising shared nodes in overlapping communities. In Table~\ref{tab:defs}, we list the key concepts of this study.

\begin{table}[tb]

\caption{Definitions of key concepts used in this study.}
\centering
\begin{tabular}{p{3.8cm} p{9.5cm}}
\toprule
\textbf{Term} & \textbf{Definition} \\
\midrule

Community &
A group of nodes that are more densely or strongly connected to each other than to the rest of the network, often identified by optimising an objective function (such as Equation~\ref{eq:community}). \\

Subcommunity &
A smaller, community contained within a larger community or nested structure. \\

Building block &
A basic structural unit formed by a distinct community or by the intersection of multiple communities. \\

Noise &
Nodes that distort or obscure the true community structure, often arising from randomness or methodological artefacts. The term "ambiguously assigned node" is used synonymously but is more frequently applied in the context of empirical network structures.\\

Shared node &
A node that appears in the intersection of two or more communities. \\

Overlapping community &
Overlapping communities consist of nodes that belong to multiple groups, with some nodes belonging to two or more communities. \\

Overlapping node &
In this study, an overlapping node is defined as a node that belongs to the intersection of overlapping communities, where the intersection does not form a distinct community. \\

Nested structure &
A structure in which a building block or subcommunity is contained within a larger community. \\
\bottomrule
\end{tabular}
\label{tab:defs}
\end{table}
\subsection{Community Detection Method Used in Demonstrations}
\label{subsec:ISM}

In our community detection model, community detection involves identifying local maximum values of the quality function (Equation 3 in~\cite{KuikkaCD}), as shown in Equation~\ref{eq:community}, calculated using the elements of the influence-spreading matrix $C(s, t)$ for $s, t = 1, \ldots, N$:

\begin{equation}
\label{eq:community}
    q = \sum_{\substack{s,t \in V \\ s \neq t}} C(s, t) + \sum_{\substack{s,t \in (G-V) \\ s \neq t}} C(s, t).
\end{equation}
In Equation~\ref{eq:community}, we represent a node influencing as $s$ (source) and a node influenced as $t$ (target). The number of nodes in the network is $N$. The elements in the matrix describe the probabilities of directed influence between any two nodes in the network. Equation~\ref{eq:community} measures the strength of the division between the two factions $V$ and $G-V$ in the network $G$. A higher value of $q$ indicates stronger cohesion, enabling comparisons between different network divisions. An advantage of our approach is its ability to identify multiple overlapping community structures without the need for a predetermined number of communities, unlike some methods currently found in the literature~\cite{Newman}.

When a local maximum is identified for a subset of nodes $V$, it indicates that $V$ and its complement $G-V$ are recognised as distinct communities. The two terms in Equation~\ref{eq:community} represent the cohesion values of these communities. This method is based on the principle that a local optimum serves as a valid solution among its neighbouring candidates. Depending on the clarity of the division between the communities, there may be nodes that are not definitively associated with either community, or moving a node from one community to another may not significantly affect the value of $q$ in Equation~\ref{eq:community}. Often, groups of these ambiguous nodes create overlapping regions between communities, which can be seen as optimal solutions when merged with one of the larger structures. Whether these intersections form separate subcommunities depends on whether they also satisfy the maximisation condition of Equation~\ref{eq:community}.

\subsection{Model for Overlapping Nodes}
\label{subsec:ONM}

In this section, we outline the scope of our study, which focuses on identifying and classifying shared nodes of overlapping communities within complex network structures. We do not describe any specific community detection methods in detail, as our approach to identifying shared or overlapping nodes is general and can incorporate any method capable of producing overlapping communities. To demonstrate this, we apply the overlapping community detection method we previously presented in~\cite{KuikkaCD}.

Our approach in identifying overlapping nodes involves two main phases. First, we detect all intersections between communities, which we refer to as building blocks. Second, we evaluate each building block to determine whether it qualifies as a subcommunity. In our community detection model, this means that it is an optimal solution of the objective function defined in~\cite{KuikkaCD}.

\begin{figure} [ht]
    \centering
    \includegraphics[width=0.95\linewidth]{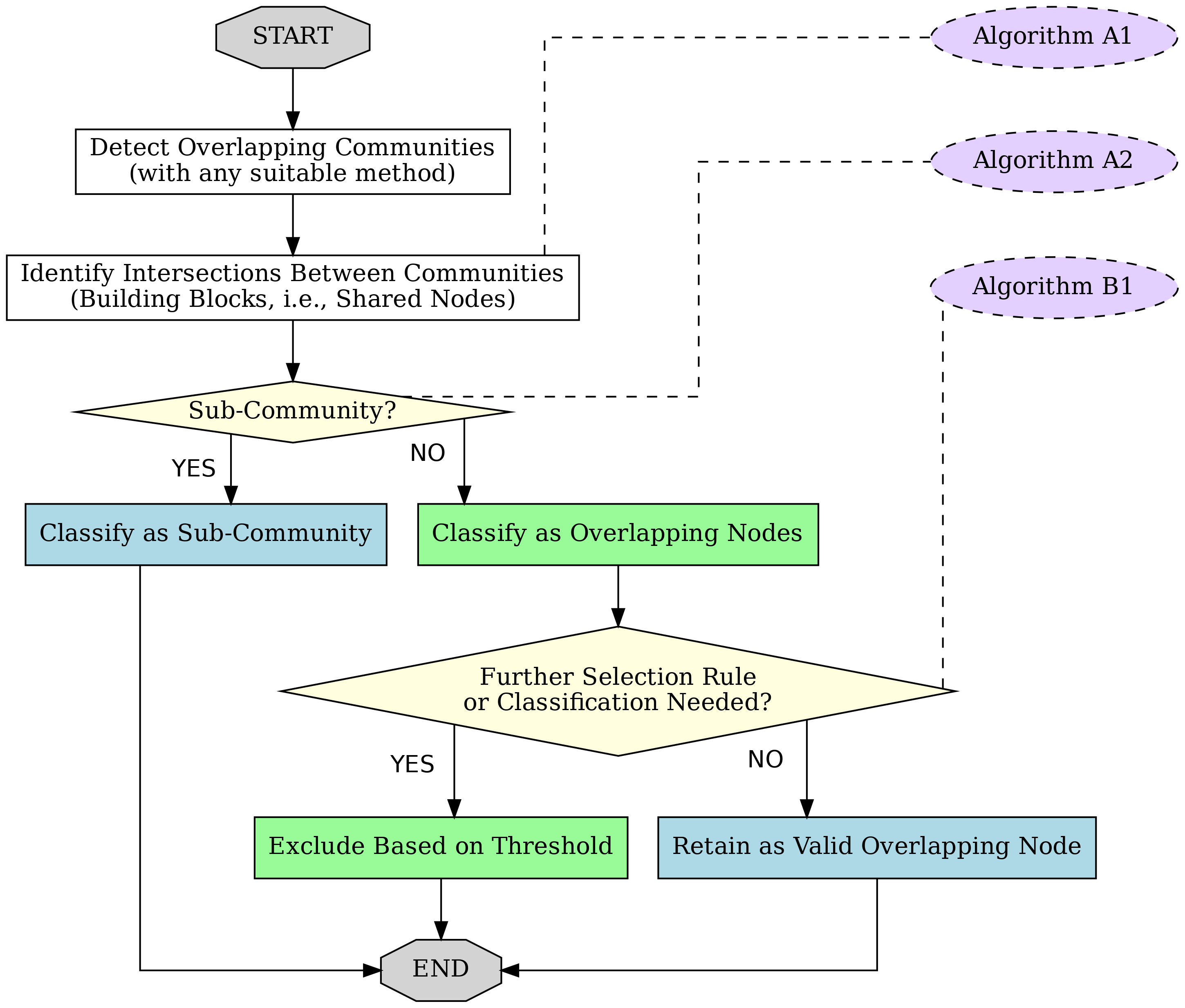}
    \caption{Diagram showing the process by which the building blocks of overlapping communities are generated and shared nodes are identified using a threshold. The purple boxes indicate the algorithms in Appendices A and B that describe each phase.}
    \label{fig:Diagram}
\end{figure}

Intersections of communities that do not qualify as individual subcommunities are candidates for overlapping nodes. In our analysis, we apply a selection rule (see Appendix~\ref{Appendix:OverlappingNodes}) to determine which nodes to retain as overlapping nodes and which to exclude. Our rule of classifying nodes as overlapping nodes is not definitive and should only be considered as an example. The flow chart in Figure~\ref{fig:Diagram} shows the main steps of the procedure. 

Our proposed technique addresses the issue of ambiguously assigned nodes and noise. The definition of noise can vary according to the specific application. In the context of community detection in complex networks, noise refers to any disturbance or uncertainty that obscures the true community structure. Noise can also arise from various sources, including data errors, random connections, incomplete sampling, intrinsic randomness in the network generation process, or limitations inherent to the community detection method itself. In our approach, we exclude noise and do not classify it as a valid overlapping region.

As the first selection criterion in our algorithm, we assess all intersections of communities to determine whether they should be recognised as separate communities. This involves checking if they are inner parts of nested communities and meet the criteria for a community, such as maximising an objective function used in community detection methods. Next, we need to define another criterion for intersections that are not categorised as subcommunities. The nodes in these intersections can all be considered overlapping nodes, but we may want to exclude some of them based on a threshold value that takes into account the number of nodes or the cohesion values~\cite{kuikka2024detailed} of the intersections. In this study, we use the number of nodes as a basis for our selection rule.

In a nested community structure, we compare the number of nodes in the outer layer with the number of nodes in the inner layer of the nested structure. As this ratio increases, the rule becomes more restrictive, resulting in more nodes being excluded and fewer nodes being counted as remaining overlapping nodes. When we increase this threshold value, only a small number of nodes, or even just a single node, may be retained as overlapping. The selection criterion indicates that, in a nested system of nearly equal-sized sets of nodes, these excluded nodes can be considered internal structures rather than overlapping nodes of different communities.

\begin{figure}[ht]
    \centering
    \begin{subfigure}{0.32\textwidth}
    \includegraphics[width=\textwidth]{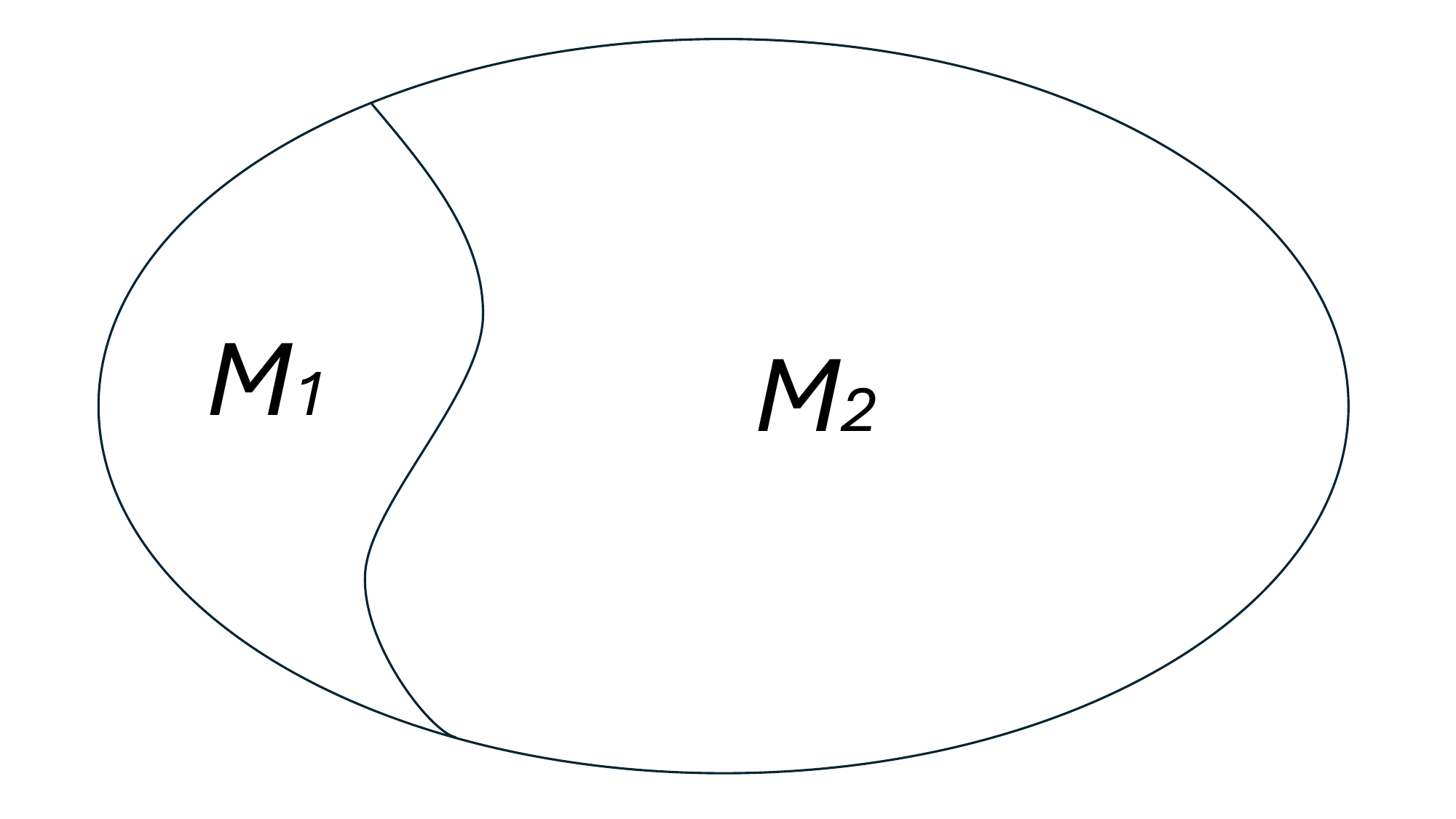}
    \captionlistentry{}
    \end{subfigure}
    \begin{subfigure}{0.32\textwidth}
    \includegraphics[width=\textwidth]{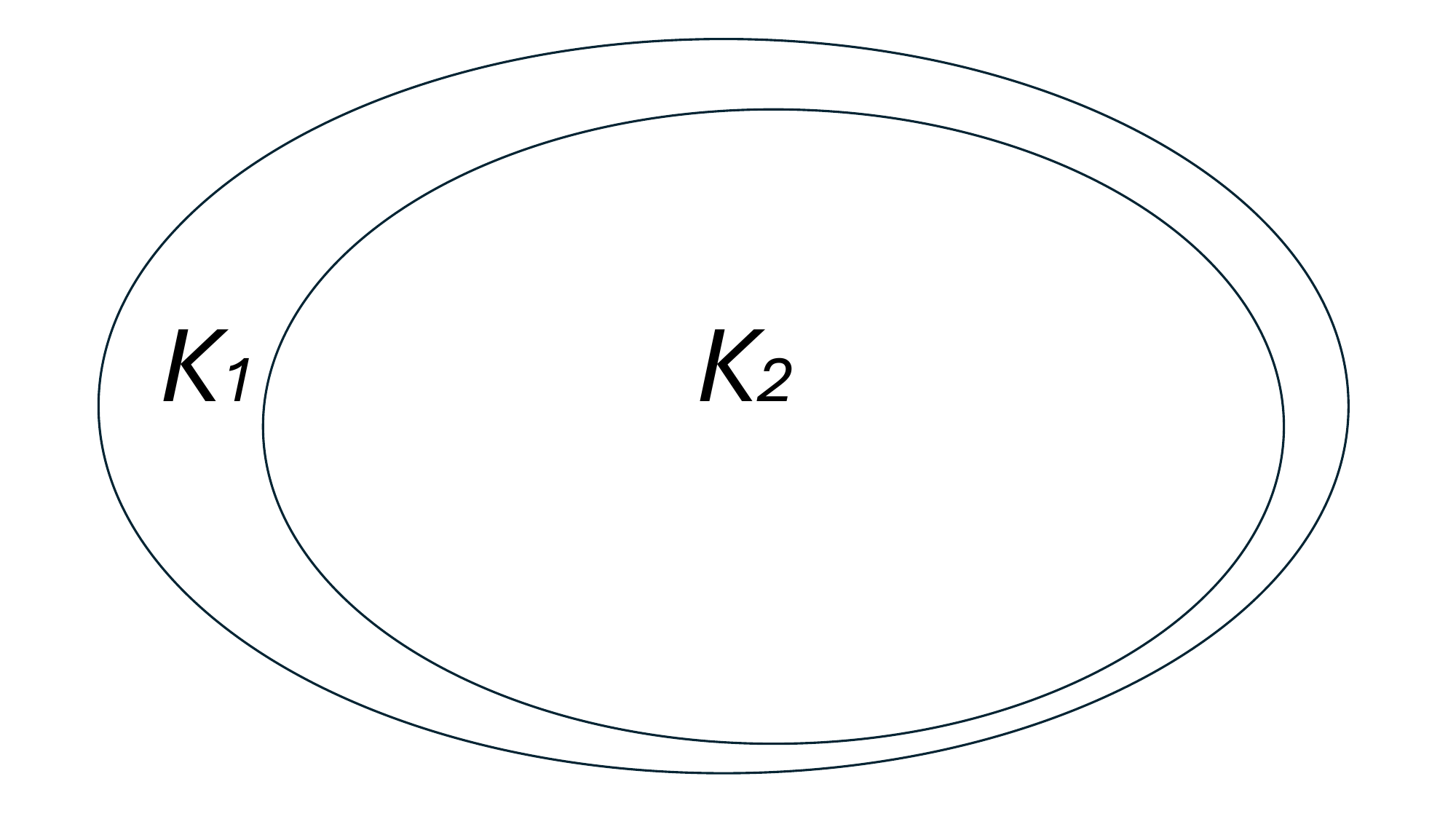}
    \captionlistentry{}
    \end{subfigure}
    \begin{subfigure}{0.32\textwidth}
    \includegraphics[width=\textwidth]{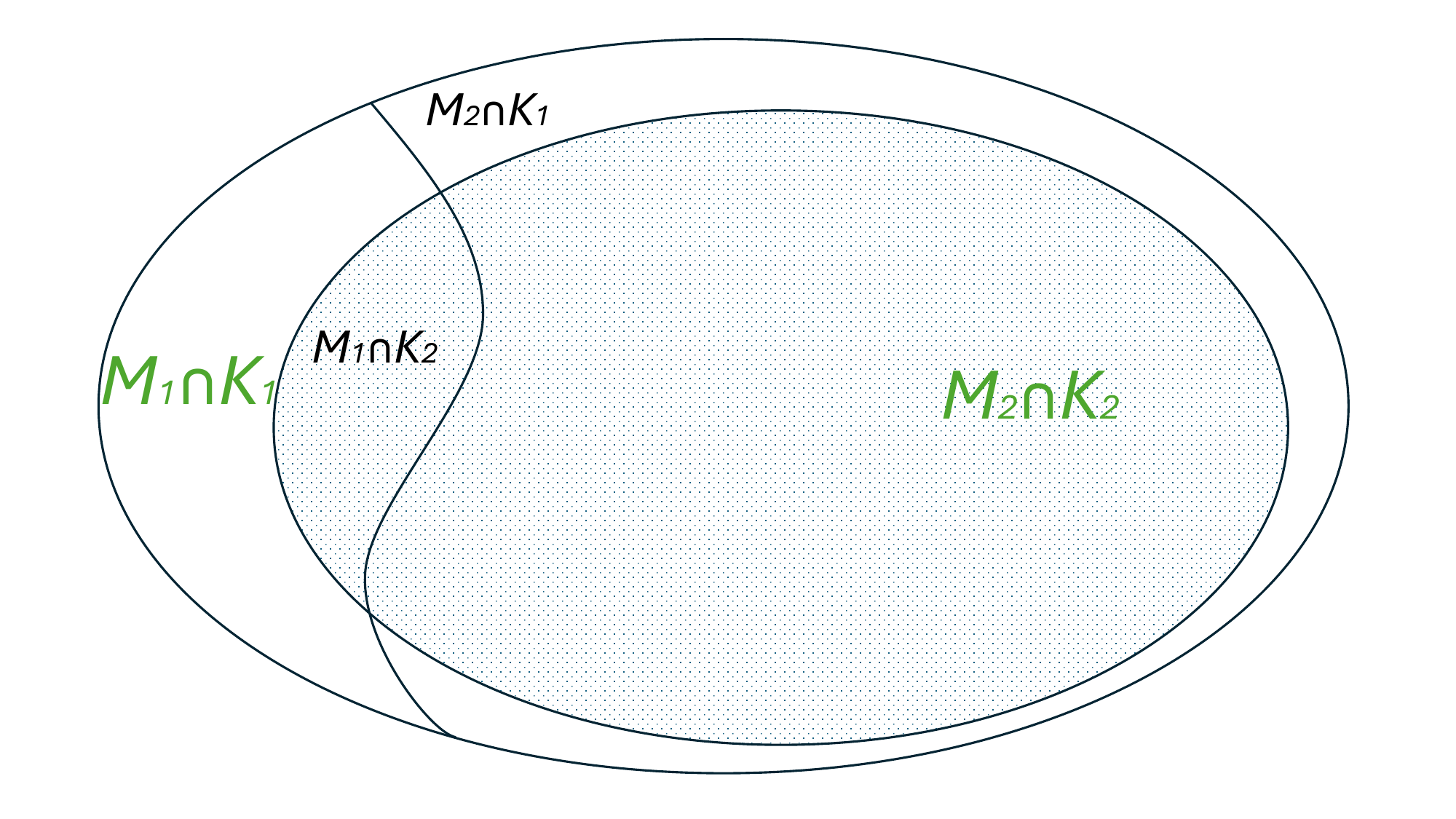}
    \captionlistentry{}
    \end{subfigure}
    \caption{An illustrative example of two divisions of the network into two communities $M_1$ and $M_2$ (left) and $K_1$ and $K_2$ (middle). Nodes in $K_2$ are almost the same as in $M_2$. The interception of these communities is marked as $M_2\cap K_2$ (right).}
    \label{fig:kuva1}
\end{figure}

Figure~\ref{fig:kuva1} illustrates a case in which community detection involves dividing the network into two segments. The figures on the left and middle show two different solutions of the community detection method. The figure on the right combines these segments and depicts various intersecting and overlapping regions: $M_1 \cap K_2$, $M_2 \cap K_2$, $M_1 \cap K_1$, and $M_2 \cap K_1$. The nodes in these regions can be used to define the inclusion and exclusion criteria for overlapping nodes. In Appendix~\ref{Appendix:OverlappingNodes}, we provide an example based on the number of nodes in the sets $K_1$ and $M_2 \cap K_2$ (or $K_2$ and $M_1 \cap K_1$). The threshold value for an overlap is defined by the condition that the ratio $r$ in Equation~\ref{eq:ratio} must be greater than a specified threshold parameter value, denoted by $thr$, in Appendix~\ref{Appendix:OverlappingNodes}.

\begin{equation}
   r = \frac{\text{number of nodes in community } K_1}{\text{number of nodes in intersection } M_2 \cap K_2} \geq thr
   \label{eq:ratio}
\end{equation}
Note that in our case, the set in $M_1 \cap K_2$ (or $M_2 \cap K_1$) is empty because the procedure in Appendix~\ref{Appendix:BB} has conducted pre-processing and classified the building blocks as valid subcommunities and other intersections of the detected communities. If this kind of pre-processing is not conducted, these sets can also be used in the selection rule.

\section{Data \& Demonstrations}
\label{sec:Demonstrations}
In this section, we demonstrate the practical use of our method for identifying overlapping nodes in overlapping communities. We evaluate the method by applying it to five real-world social networks, shown in Table~\ref{tab:Networks}. Four of the networks are drawn from the prior literature, while the dataset in Section~\ref{subsec:Mobile} is derived from a large mobile phone call detail record (CDR) database \cite{Onnela2007}. Data are represented as social networks in which nodes denote individuals, and edges represent social ties between them. Next, we describe the datasets and their key characteristics, followed by the corresponding results. For each network, we compare the empirically observed community structures with the theoretical structures suggested by the network topology. For example, both Zachary’s Karate Club and the Dolphin social network are known to divide into two main groups, providing useful reference points to evaluate the results of our method. However, it should be noted that these ground truths are not necessarily definitive, and thus it is not possible to determine with certainty whether our method successfully converges towards the community structures agreed in the literature. In the case of Facebook, the comparison against a ground truth is particularly challenging because the labels are manually assigned and may themselves contain inconsistencies or subjective judgments. For the mobile phone network, no ground truth is available due to the private and anonymous nature of the data, making direct validation infeasible. For Zachary's Karate club and Dolphin social networks, we provide tables to compare our approach against selected studies in the literature.
\begin{table}[h]
    \centering
    \caption{Five social networks used in our demonstrations.}
    \begin{tabular}{|c|c|c|c|c|c|}\hline
 Section & Network & Nodes & Edges & $\langle C \rangle$
  & Threshold \\
 &&&&&values\\\hline\hline
 \ref{subsec:Karate} & Zachary's Karate Club & 34 & 78 & 0.57 & 0.0, 0.5\\\hline
 \ref{subsec:Kurjat} & Les Misérables & 77 & 254 & 0.57 & 0.0, 1.0, 3.0\\\hline
 \ref{subsec:Mobile} & Mobile Phone Call Data & 4270 & 4766 & 0.05 & 0.0, 0.075, 0.1, 0.2\\\hline
 \ref{subsec:Dolphins} & Dolphins & 66 & 159 & 0.26 & 0.0, 1.0, 5.0\\\hline
 \ref{subsec:Facebook} & Facebook Social Circles & 4039 & 88234 & 0.61 & 0.0, 1.0, 2.0\\
 &&&&&10.0, 20.0, 60.0 \\\hline
    \end{tabular}
    \label{tab:Networks}
\end{table}

In Table~\ref{tab:Networks}, we present the network statistics, including the number of nodes and edges, the average clustering coefficient $\langle C \rangle$, and the values of the corresponding threshold parameters used in our analysis. These threshold values have been specifically chosen to highlight the significant changes in the set of overlapping nodes as the threshold increases above zero. It is important to note that these values are influenced by the results of the overlapping community detection methods and the characteristics of the studied network. Consequently, there is no universally applicable set of threshold values.

\subsection{Zachary's Karate Club Social Network}
\label{subsec:Karate}

\begin{figure}[ht]
    \centering
    \begin{subfigure}{0.49\textwidth}
    \includegraphics[width=\textwidth]{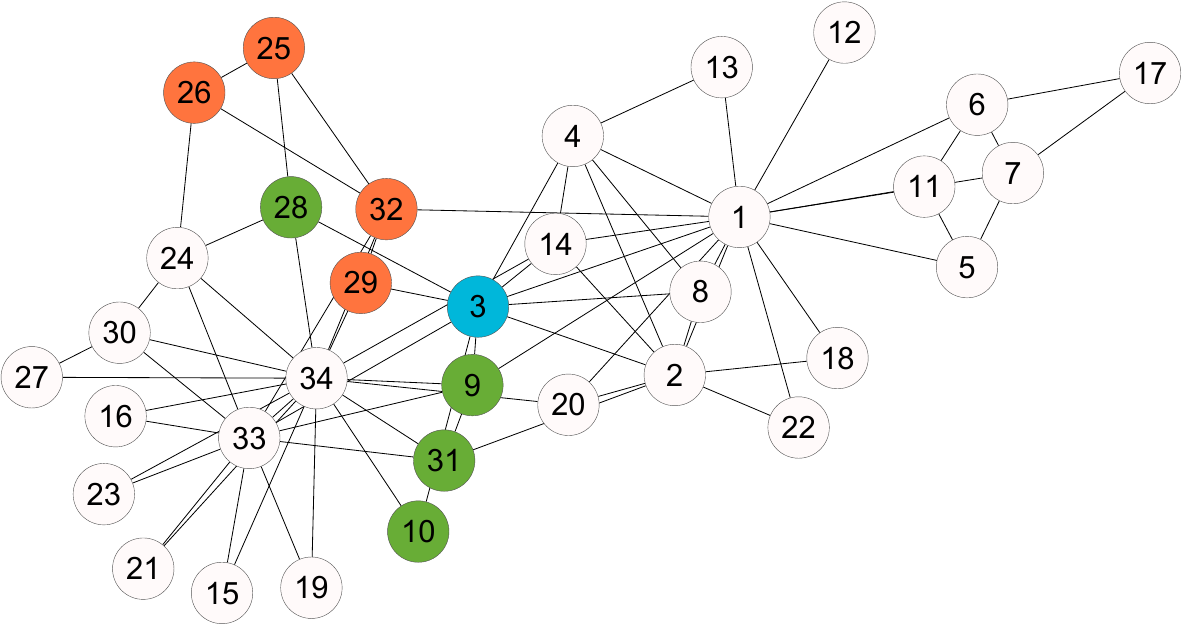}
    \captionlistentry{}
    \label{fig:Za00_lo}
    \vspace{2 mm}
    \end{subfigure}
    \begin{subfigure}{0.49\textwidth}
    \includegraphics[width=\textwidth]{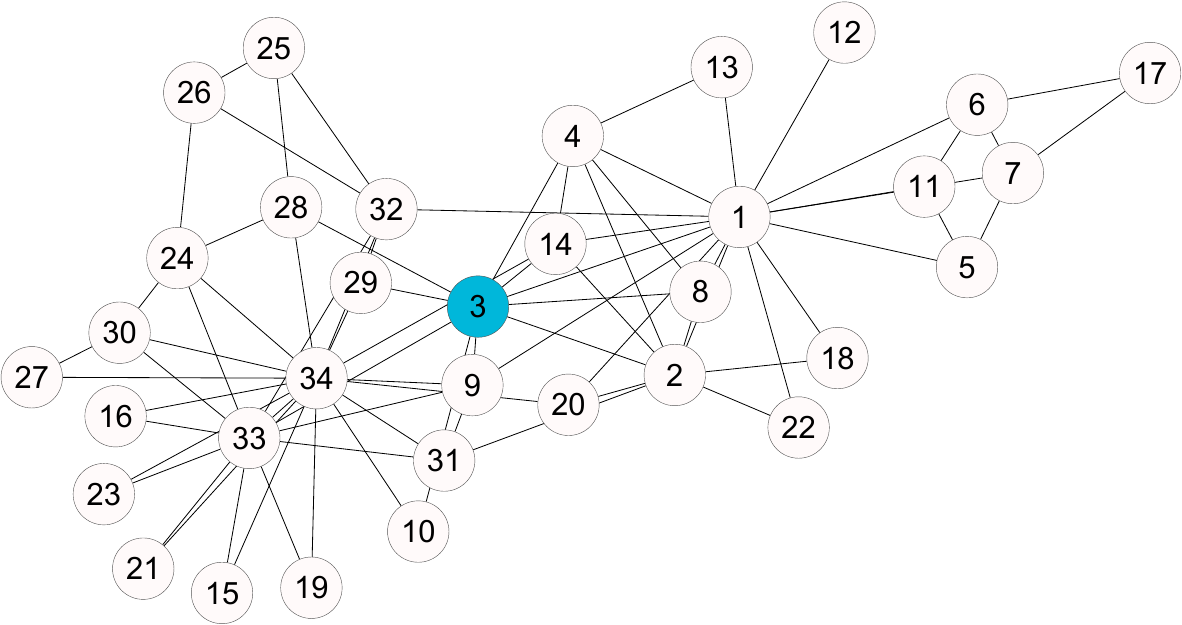}
    \captionlistentry{}
    \vspace{2 mm}
    \label{fig:Za10_lo}
    \end{subfigure}
    \caption{Zachary's Karate Club network overlapping nodes identified for two threshold parameter values ($thr=0.0$ and $0.5$).}
    \label{fig:Zac005}
\end{figure}

In the Zachary's Karate Club network (Figure~\ref{fig:Zac005}), which illustrates friendships between $34$ club members, a dispute between the instructor (node $1$) and the administrator (node $34$) resulted in the club splitting into two factions that closely reflected the structure of the network. Both leaders were highly central nodes with numerous connections, and members generally sided with the leader to whom they were more closely connected. Node $3$, who had strong ties to both leaders, eventually joined the instructor's faction, according to Zachary's analysis~\cite{Zachary1977}. The only member whose choice did not align with the prediction of the model was node $9$, who sided with the administrator instead of the instructor. This case highlights how the structure of social networks can effectively predict real-world group divisions.

Several studies have applied different methods to detect overlapping communities in the Zachary Karate Club network~\cite{Newman2006Modularity} and found that although the original division was clearly binary, certain nodes consistently appear at the boundary between communities. Across different algorithms, nodes $3, 9, 10, 14,$ and $31$ are most frequently identified as overlapping, indicating their connections to members of both factions. For example, the study in~\cite{Chen2014SelfCorrecting} discovered that nodes $3, 9,$ and $31$ belong to multiple groups according to the self-correcting algorithm. Similarly, the study in~\cite{Zhang2018VitalNodes} identified the nodes $3, 9, 14,$ and $31$ as the boundary nodes. The study in~\cite{Dehghan2021TES} reported nodes $3$ and $10$, as well as node $10$ separately, as overlapping nodes. Furthermore, the study in~\cite{Duan2019MultiType} characterised node $9$ as an overlapping node. These results consistently indicate that a small group of individuals, particularly the nodes $3, 9, 10,$ and $31$, act from a structural point of view as bridges or connectors between the two main communities. This shows how certain members’ social ties can blur strict community boundaries within the club. The comparison of nodes identified as overlapping in the literature is summarised in Table \ref{tab:overlapping_nodes}. These examples represent only a small subset of the available studies and are not intended to be exhaustive.

\begin{table}[ht]
\centering
\caption{Comparison of overlapping nodes in the Zachary's Karate Club identified across studies.}
\begin{tabular}{lcccccccccccc}
\multicolumn{2}{c}{} & \multicolumn{10}{c}{\textbf{Nodes}} \\
\hline
\textbf{Study} & \textbf{Source} & \textbf{3} & \textbf{9} & \textbf{10} & \textbf{14} & \textbf{25} & \textbf{26} & \textbf{28} & \textbf{29} & \textbf{31} & \textbf{32} \\
\hline
Chen et al. (2014) & ~\cite{Chen2014SelfCorrecting} & $\times$ & $\times$ &  &  &  &  &  &  & $\times$ &  \\
Zhang et al. (2018) & ~\cite{Zhang2018VitalNodes} & $\times$ & $\times$ &  & $\times$ &  &  &  &  & $\times$ &  \\
Dehghan et al. (2021) & ~\cite{Dehghan2021TES} & $\times$ &  & $\times$ &   &  &   &  &   & &  \\
Duan et al. (2019) & ~\cite{Duan2019MultiType} &  & $\times$ &  &  &  &  &  &  &  &  \\
\textbf{Our approach ($thr=0.0$)} & ~\cite{KuikkaCD} & $\times$ & $\times$ & $\times$ &  & $\times$ & $\times$ & $\times$ & $\times$ & $\times$ & $\times$ \\
\hline
\end{tabular}
\label{tab:overlapping_nodes}
\end{table}

Next, we compare the results of our model with the results of other theoretical models. Nodes $9, 10,$ and $31$ are identified as overlapping nodes, consistent with the theoretical models mentioned above. Furthermore, node $28$ is also recognised as an overlap node. Nodes $9, 10, 28,$ and $31$ exist consistently within the same subcommunity, as indicated in~\cite{KuikkaCD}. When they change subcommunities, they do so together, which is represented by the dark green colour in Figure~\ref{fig:Zac005}. Similarly, nodes $25, 26, 29,$ and $32$ form a similar group, but are typically not identified as overlapping nodes in the existing literature. This is because our community detection method~\cite{KuikkaCD} identifies more subcommunities. However, these four nodes are classified as one of the four communities in the study~\cite{Dehghan2021TES}. This highlights the challenge of making clear distinctions between communities and the potential for overlap among communities in different models. In our model, node $3$ remains classified as an overlapping node even for high values of the threshold parameter, which is consistent with other theoretical findings in the literature. In contrast to the other theoretical models, node $14$ is not considered to overlap in our model. Figure~\ref{fig:Zac005} illustrates the overall situation in which overlapping nodes are distributed between the peripheral parts of the network.

\subsection{Les Misérables Social Network}
\label{subsec:Kurjat}

Several studies have used the Les Misérables social network, which consists of $77$ fictitious characters from Victor Hugo's book, as a benchmark for detecting overlapping or mixed-membership communities~\cite{Knuth1993,Kunegis2013KONECT,riolo}. network illustrates character co-occurrences, such that each node represents a character and each edge connects two characters who appear together in a scene or chapter. The social network within the narrative features several distinct groups: Valjean's group (nodes $12, 24$, and $27$), the revolutionary students (nodes $54-65$), and the Thenardier family (nodes $25, 26, 29$, and $30$). Together, these groups consist of $17$ nodes out of the total of $77$ characters in the book who engage in social interactions and contribute to the community structure.

\begin{figure}[ht]
    \centering
    \begin{subfigure}{0.32\textwidth}
    \includegraphics[width=\textwidth]{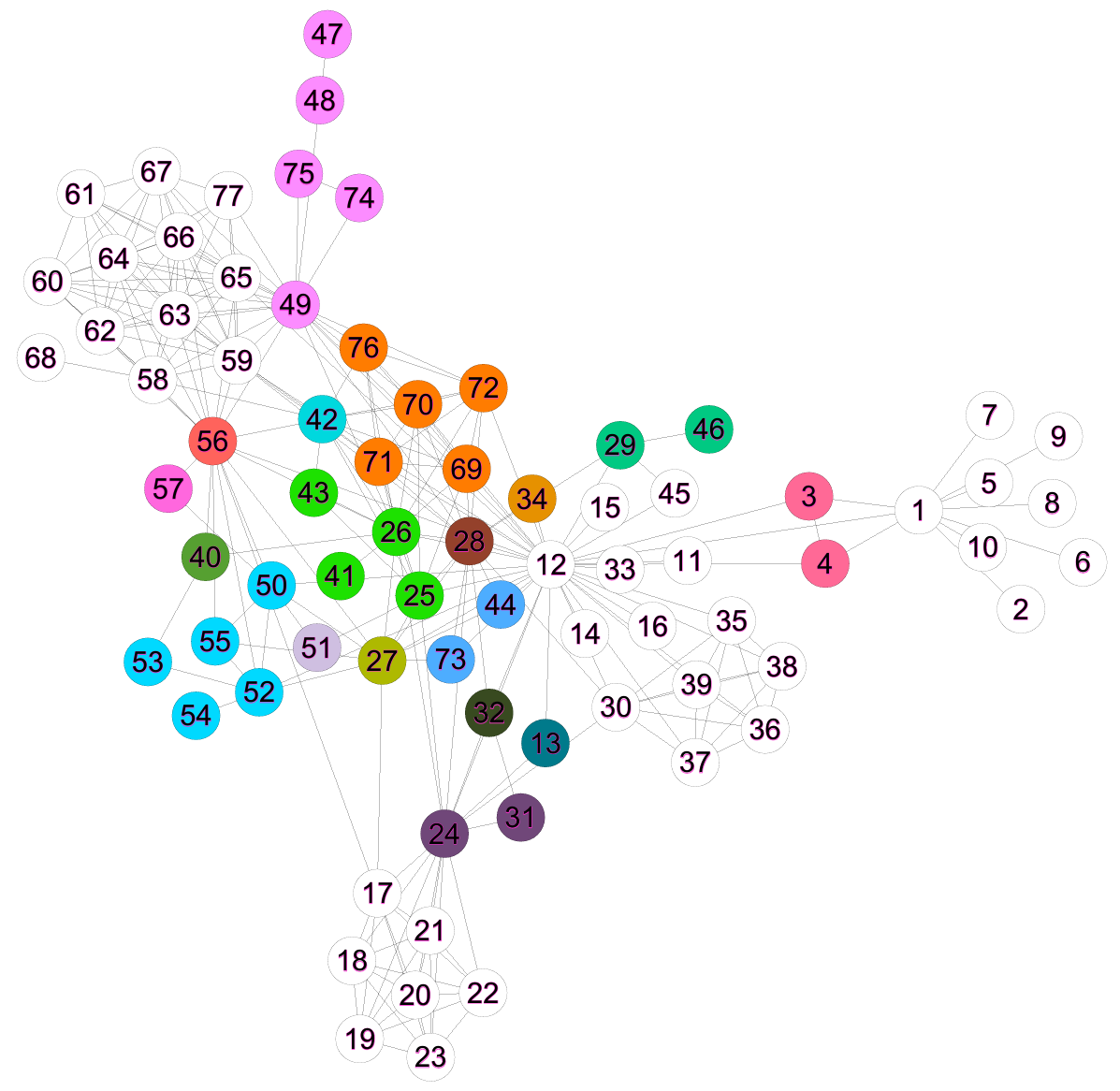}
    \captionlistentry{}
    \label{fig:les000}
    \vspace{2 mm}
    \end{subfigure}
    \begin{subfigure}{0.32\textwidth}
    \includegraphics[width=\textwidth]{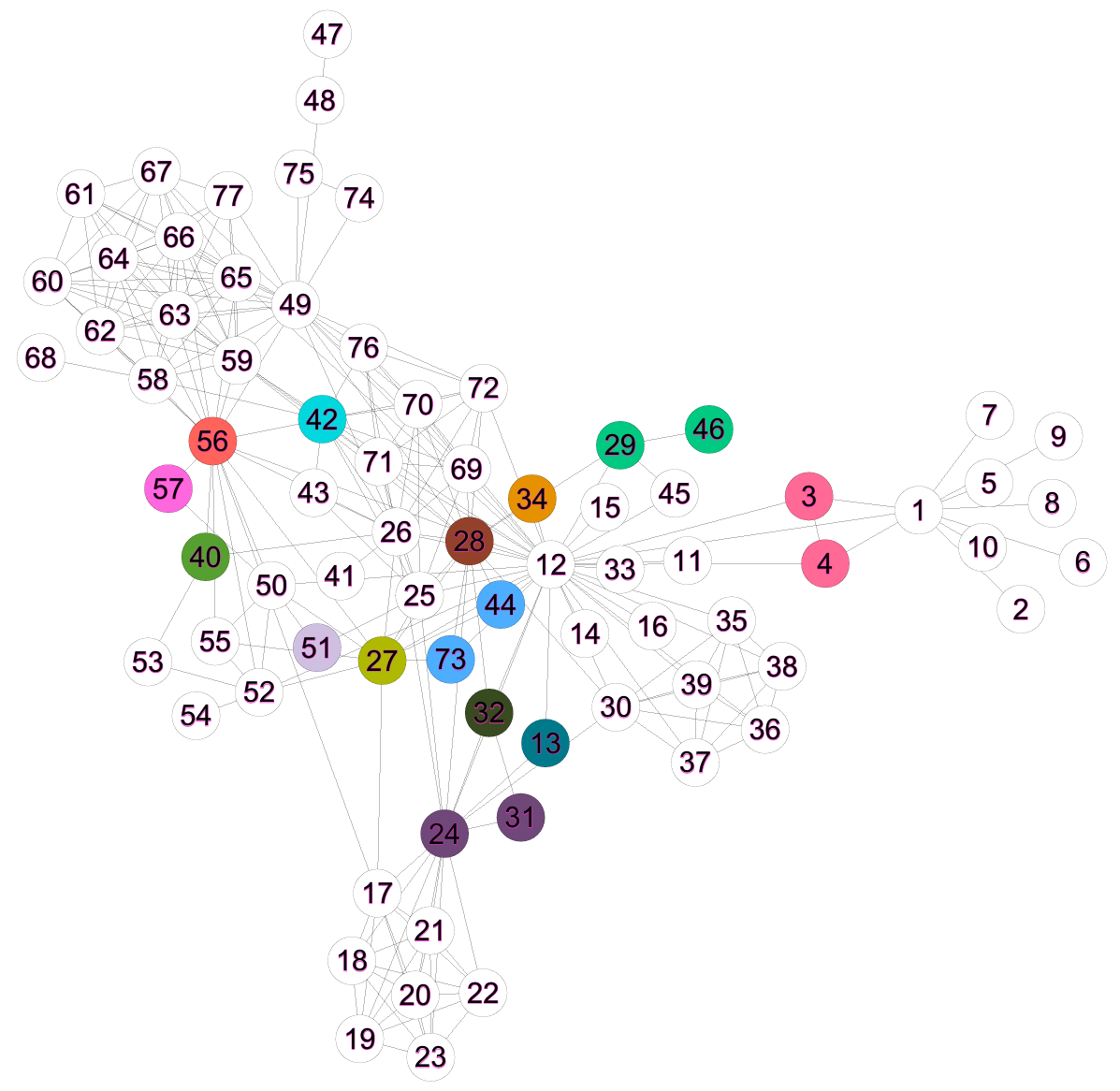}
    \captionlistentry{}
    \vspace{2 mm}
    \label{fig:les100}
    \end{subfigure}
    \begin{subfigure}{0.32\textwidth}
    \includegraphics[width=\textwidth]{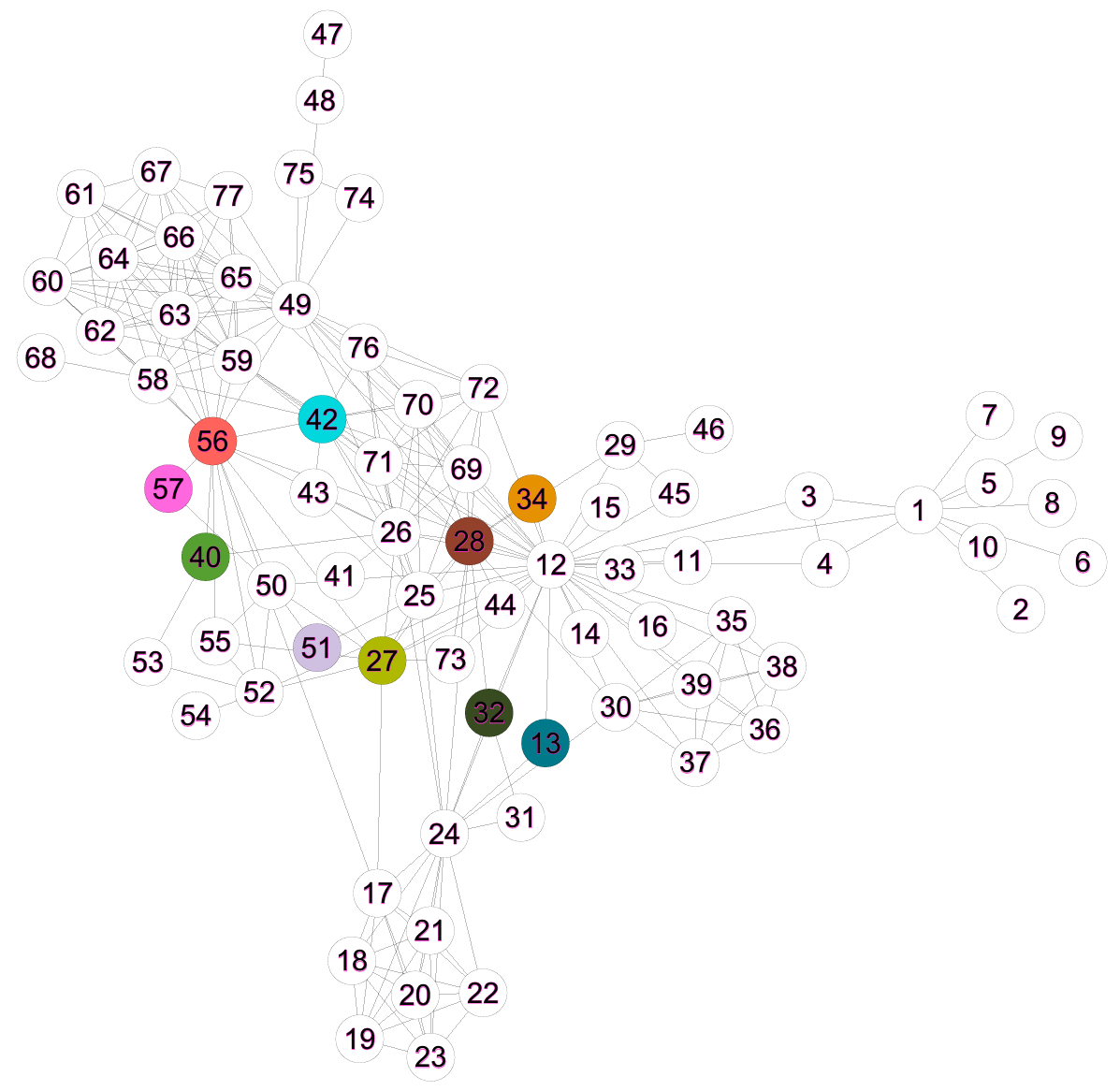}
    \captionlistentry{}
    \vspace{2 mm}
    \label{fig:les300}
    \end{subfigure}
    \caption{Les Misérables network overlapping nodes identified for three increasing threshold parameter values ($thr=0.0,1.0$, and $3.0$).}
    \label{fig:Les005}
\end{figure}

The community structure does not strictly reflect these groups, except perhaps for the revolutionary group, because the story revolves primarily around the social interactions between various characters. The communities and their overlapping nodes encompass all the families and other individuals, highlighting the author's broader storyline that extends beyond mere interactions among these families. Some characters in the narrative function as overlapping or boundary nodes, serving as bridges between different communities and connecting various social groups. In Figure~\ref{fig:Les005}, the overlapping nodes are highlighted in different colours, illustrating the relationships that reflect the author's story development in the book.

\begin{figure}[ht]
    \centering
    \begin{subfigure}{0.32\textwidth}
    \includegraphics[width=\textwidth]{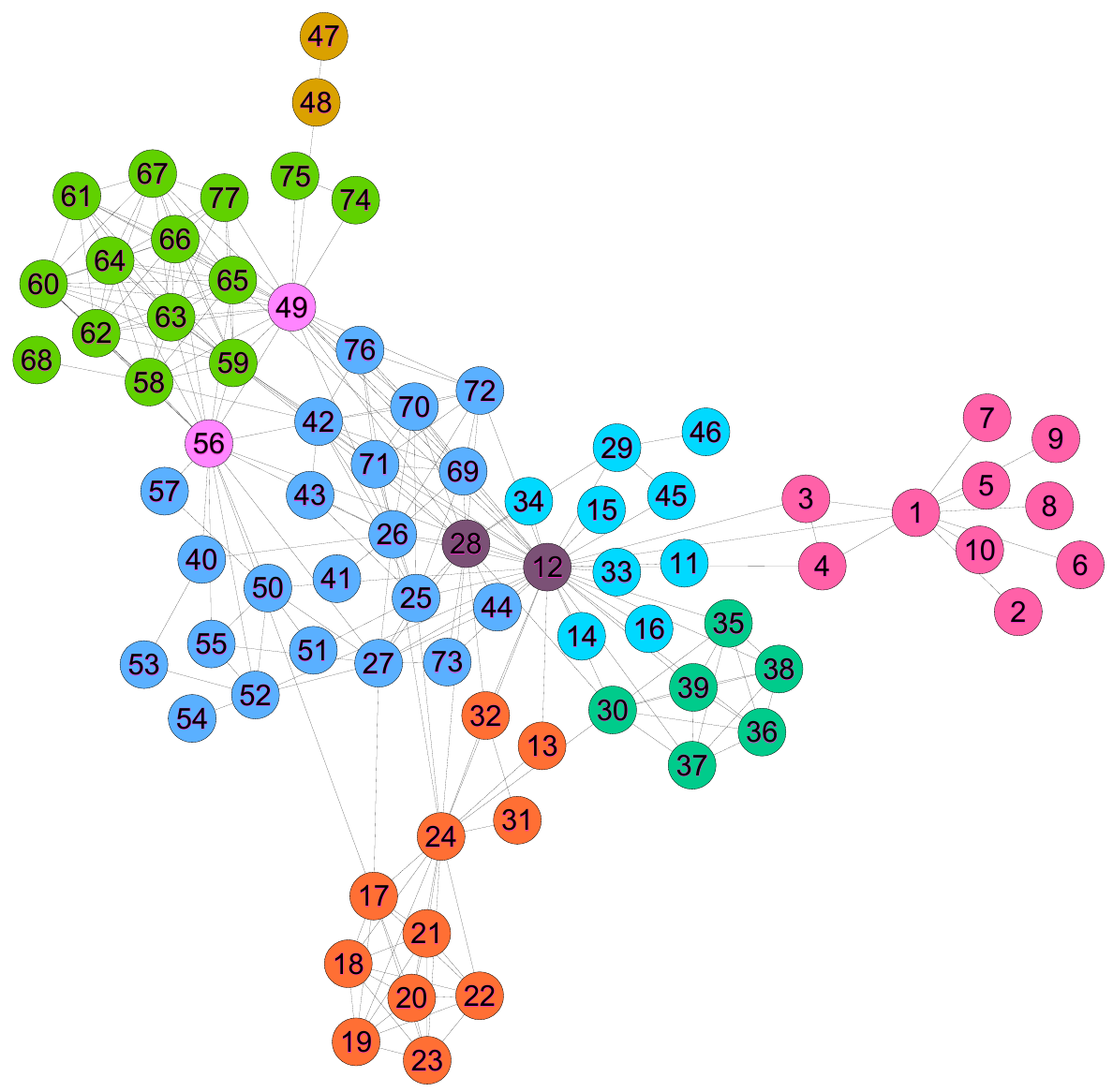}
    \captionlistentry{}
    \label{fig:RioloBuildingBlocks}
    \vspace{2 mm}
    \end{subfigure}
    \caption{Building blocks identified by Riolo and Newman using the information-theoretic method in their study~\cite{riolo}. These building blocks include both communities and their intersections.}
    \label{fig:Riolo}
\end{figure}

To study the community structure of the Les Misérables social network, Riola and Newman used an information-theoretic approach~\cite{riolo}. Applying techniques such as information compression-based network partitioning, they identified “building blocks” of communities—groups of characters that frequently interact and form cohesive substructures within the overall network, as shown in Figure~\ref{fig:Riolo}. The structure of the eight blocks at the mutual information peak includes both communities and their intersections. We compare this structure with our first figure, which uses the parameter value of $thr=0.0$, as shown in Figure~\ref{fig:Les005}. In our model, the white nodes are identified as members of the subcommunities or their internal structure, and the coloured nodes represent the overlapping nodes for different threshold values. The study in~\cite{riolo} demonstrated how information theory can reveal the hierarchical and modular organisation of complex social systems.

As a result of this comparison, we conclude that although there are minor differences, the largest building block, indicated by the dark blue colour in Figure~\ref{fig:Riolo}, corresponds to the overlapping nodes in Figure~\ref{fig:Les005}. The other five larger blocks, represented by shades of light green, dark green, light blue, orange, and purple, correspond to distinct subcommunities in our model.

\subsection{Empirical Mobile Phone Call Network}
\label{subsec:Mobile}

The empirical dataset analysed in this section is based on data collected from a large mobile phone call detail records (CDRs) database. This dataset has been used to study human communication patterns, its social structure, and behavioural dynamics~\cite{Li2014MotifsMobile}. Derived from large-scale anonymised call detail records that cover millions of interactions, the data capture ego-centric social networks where nodes represent individuals and weighted edges indicate intensity or frequency of communication. Research based on this dataset has explored a variety of topics, including the strength of social links and gender- or age-related differences in communication patterns~\cite{Fudolig2020SocialCloseness}, mobility and internal migration in social interactions~\cite{Fudolig2022MigrationCommunication}, and the evolution of statistically validated motifs and temporal communication structures. Collectively, these studies demonstrate how large-scale mobile phone data reveal consistent behavioural regularities, such as the persistence of close social bonds, demographic trends in calling behaviour, and stable network patterns, while offering valuable insights into human social organisation and communication dynamics at multiple scales.

\begin{figure}[ht]
    \centering
    \begin{subfigure}{0.45\textwidth}
    \includegraphics[width=\textwidth]{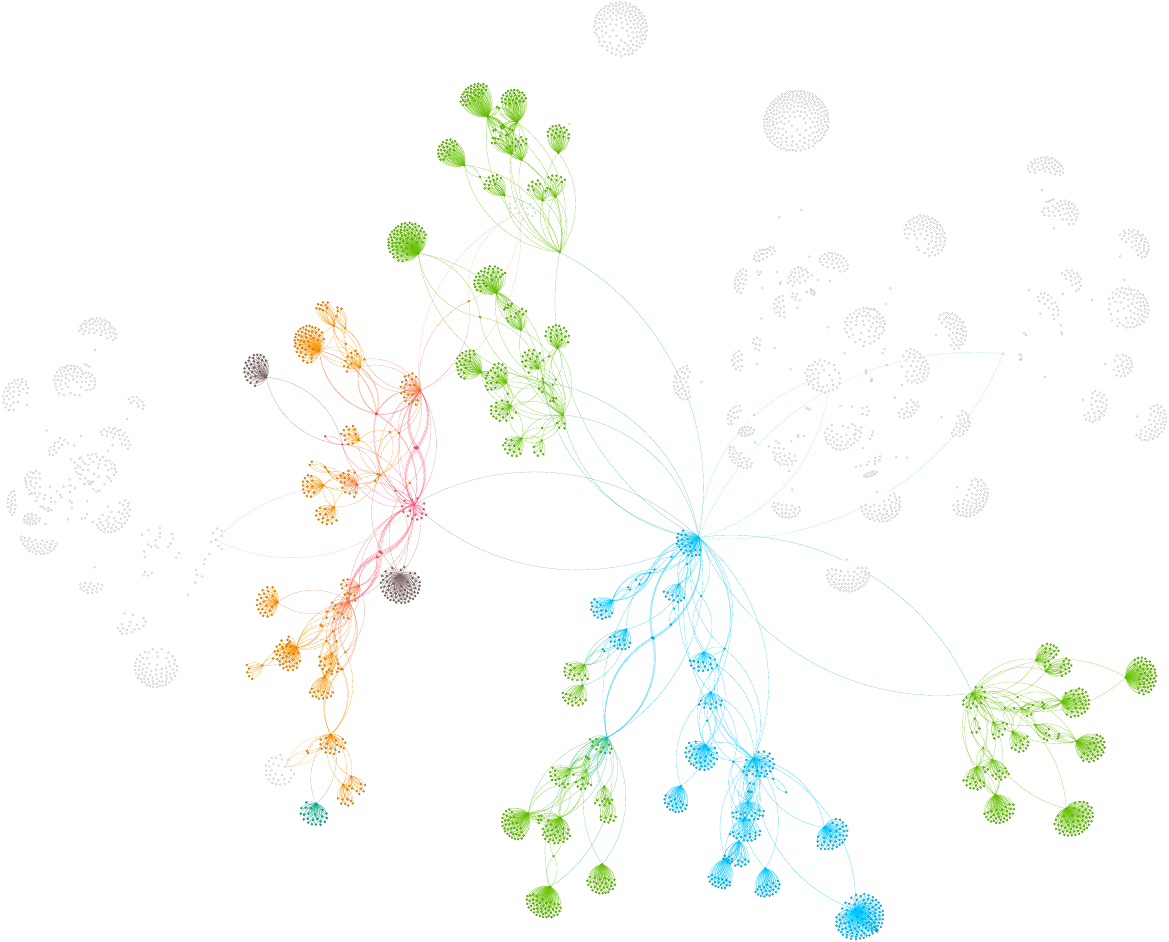}
    \captionlistentry{}
    \label{fig:bmo_lo_00}
    \vspace{2 mm}
    \end{subfigure}
    \hspace{5 mm}
    \begin{subfigure}{0.45\textwidth}
    \includegraphics[width=\textwidth]{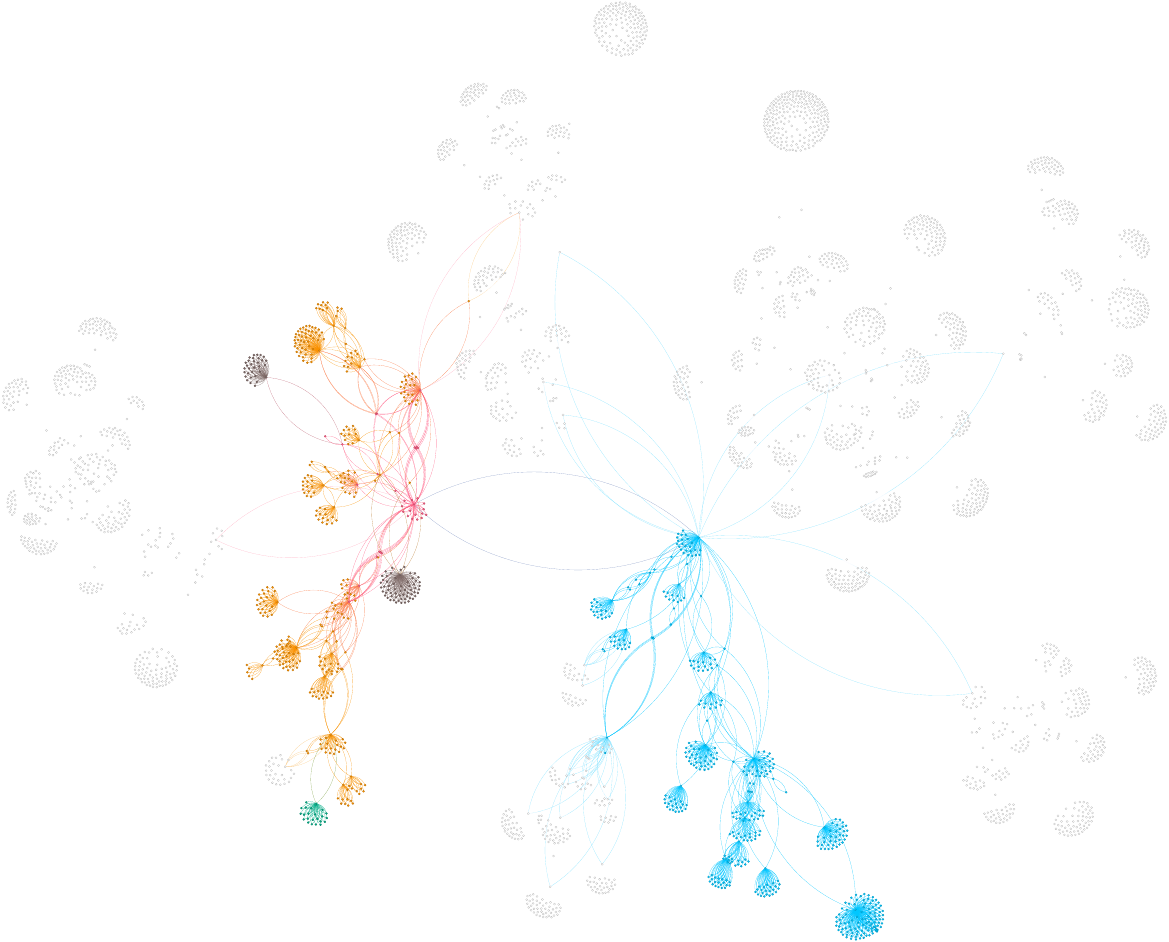}
    \captionlistentry{}
    \vspace{2 mm}
    \label{fig:bmo_lo_0075}
    \end{subfigure}
    \begin{subfigure}{0.45\textwidth}
    \includegraphics[width=\textwidth]{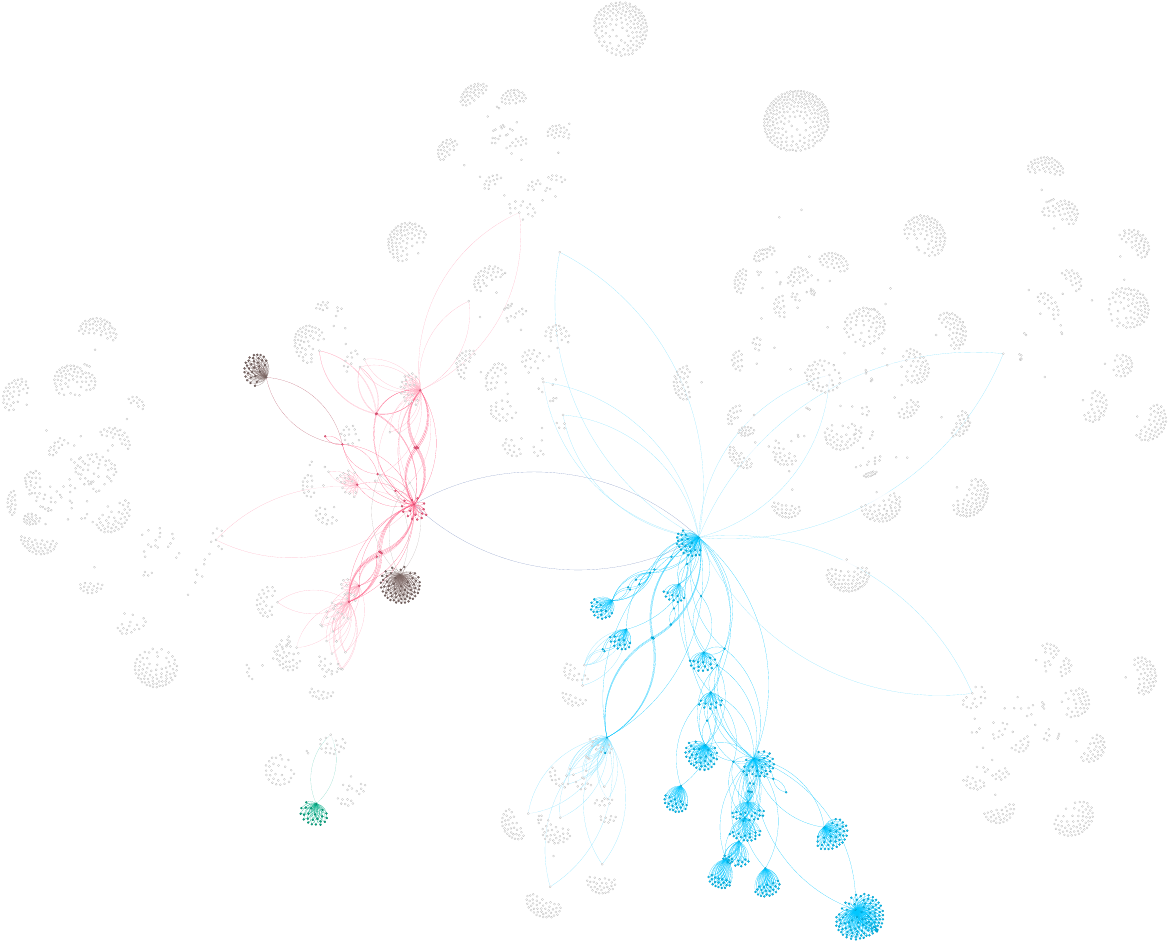}
    \captionlistentry{}
    \label{fig:bmo_lo_01}
    \vspace{2 mm}
    \end{subfigure}
    \hspace{5 mm}
    \begin{subfigure}{0.45\textwidth}
    \includegraphics[width=\textwidth]{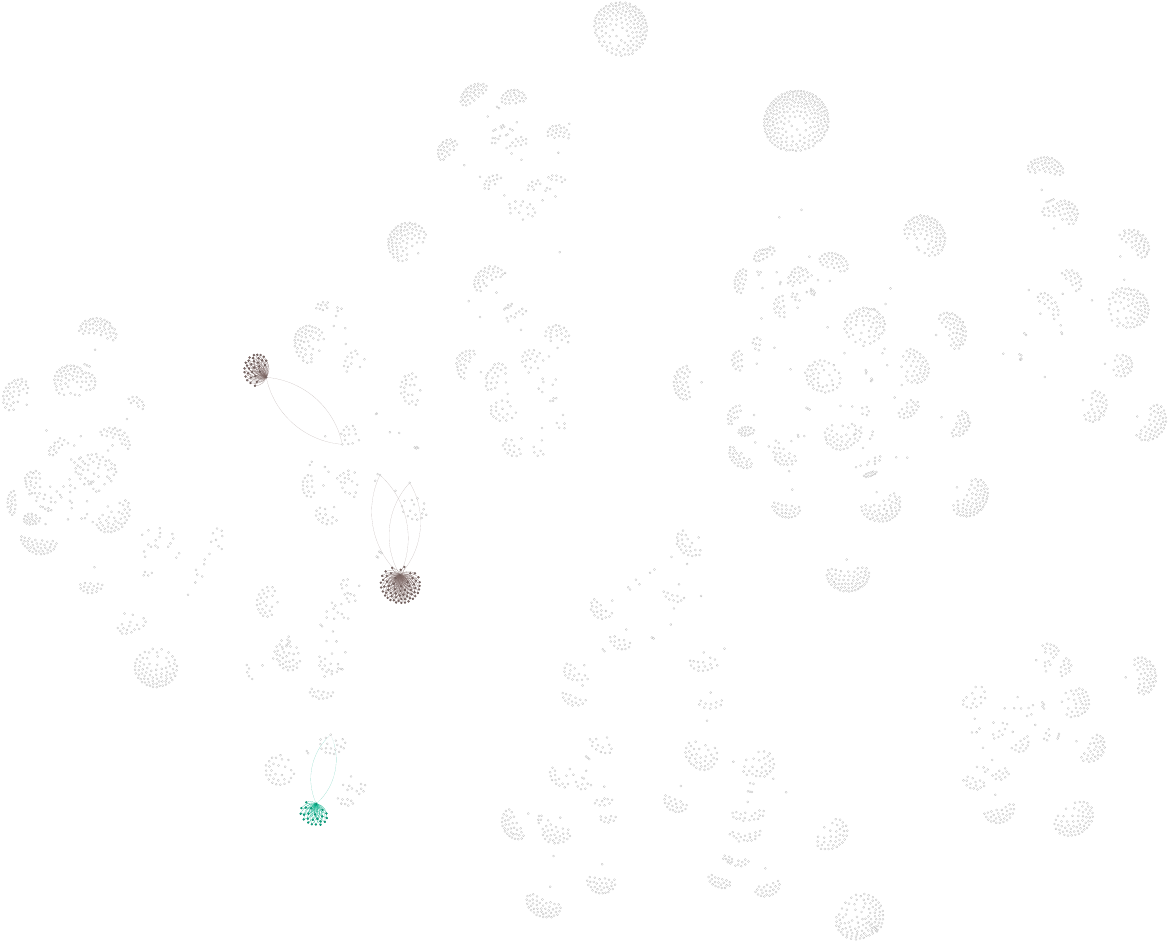}
    \captionlistentry{}
    \label{fig:bmo_lo_02}
    \vspace{2 mm}
    \end{subfigure}
    \caption{An example of overlapping nodes in a mobile phone call network, shown for four progressively increasing threshold values ($thr=0.0, 0.075, 0.1,$ and $0.2$).}
    \label{fig:bmo}
\end{figure}

Here, we focus solely on the topological properties of a subset of the mobile phone call network. The four panels in Figure~\ref{fig:bmo} that show the effects of increasing the value of the threshold parameter of our model reveal a pattern that is consistent with other cases in this study. We observe that the number of overlapping nodes decreases as the value of the threshold parameter increases. In the fourth panel, only three small groups remain as overlapping. Compared to the other networks examined, the network is sparse: communities are clearly separated and typically connected by only one inter-community edge. This results in disjoint building blocks even when the threshold parameter is set to zero. As expected, a small increase in the threshold further reduces the remaining overlaps. In certain applications, even these few remaining overlapping groups can be further analysed to determine the most central focal node of them. An example of this is the problem of optimal sensor placement in a communication network, where it is feasible to have only one sensor for each of these groups.

\subsection{Dolphin Social Network}
\label{subsec:Dolphins}

The following example highlights an undirected social network of $62$ bottlenose dolphins in Doubtful Sound, New Zealand, between 1994 and 2001~\cite{Lusseau}. In this network, nodes represent individual dolphins, while edges connect pairs observed together more often than expected by chance, indicating significant social associations. Analysed by Lusseau and Newman~\cite{NewmanLusseau}, the dataset reveals a cohesive, yet modular, community structure.

During this study, one dolphin, SN100, temporarily disappeared, causing the network to split into two subgroups~\cite{NewmanLusseau}. When SN100 returned, the groups reunited, demonstrating how the presence or absence of a central individual can affect social cohesion. This dataset has become a key example in social network analysis, particularly for studying community structure and the influence of individuals on group dynamics.

Several studies have applied mixed-membership and overlapping community detection methods to the Dolphin social network. Although the network naturally divides into two main social groups, certain dolphins consistently appear at the boundary between communities, suggesting overlapping social affiliations. For example, the study in~\cite{Shen2009Overlap} identified overlapping individuals that included Beak, Kringel, MN105, Oscar, PL, SN4, SN9, and TR99, while the study in~\cite{Gopalan2017MixedMembership} found Beak, Bumper, Fish, Oscar, PL, SN89, SN96, and TR77 as boundary dolphins under a mixed membership model. The original analysis by Lusseau and Newman in~\cite{Lusseau} highlighted SN100 as the key bridge individual on the dolphin social network. Together, these results illustrate that certain dolphins, particularly Beak, Oscar, PL, and SN100, play crucial connector roles between subgroups, emphasising the importance of overlapping and bridging individuals in maintaining social cohesion. Next, we compare the results of our model with those previously mentioned.

\begin{figure}[H]
    \centering
    \begin{subfigure}{\textwidth}
    \includegraphics[width=1\textwidth]{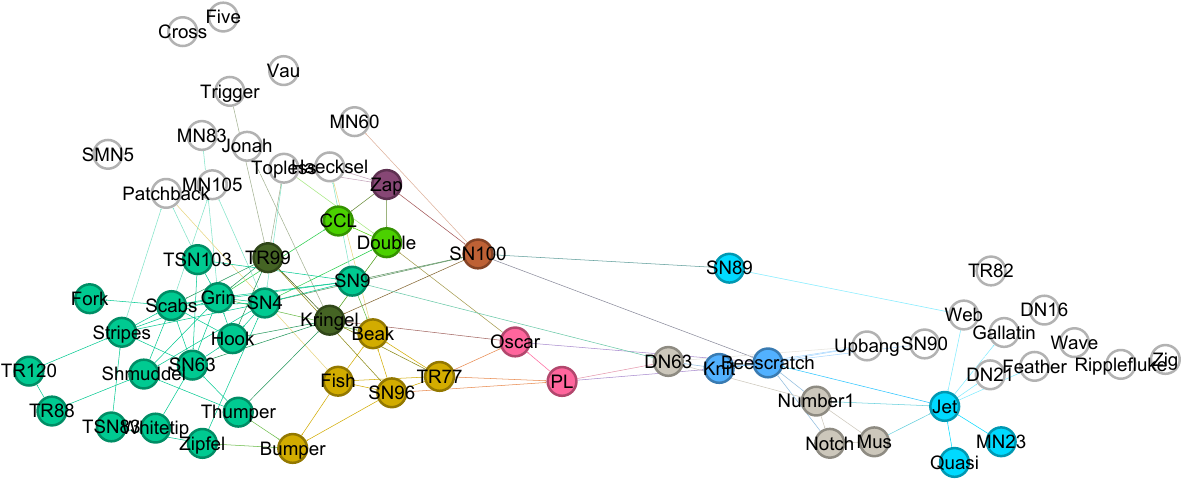}
    \vspace{2 mm}
    \label{fig:Do00_lo}
    \end{subfigure}
    \begin{subfigure}{\textwidth}
    \includegraphics[width=1\textwidth]{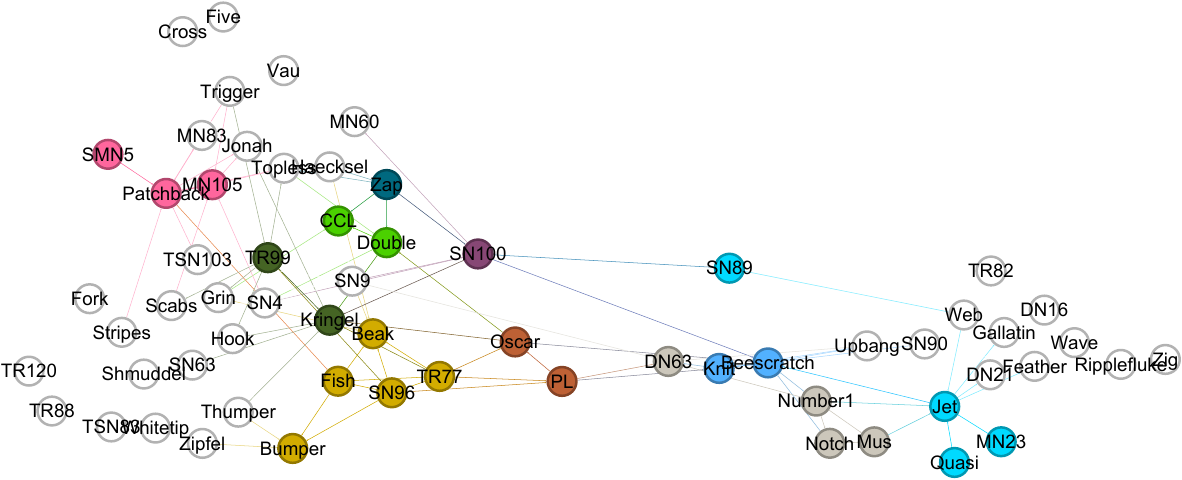}
    \vspace{2 mm}
    \label{fig:Do10_lo}
    \end{subfigure}
    \begin{subfigure}{\textwidth}
    \includegraphics[width=1\textwidth]{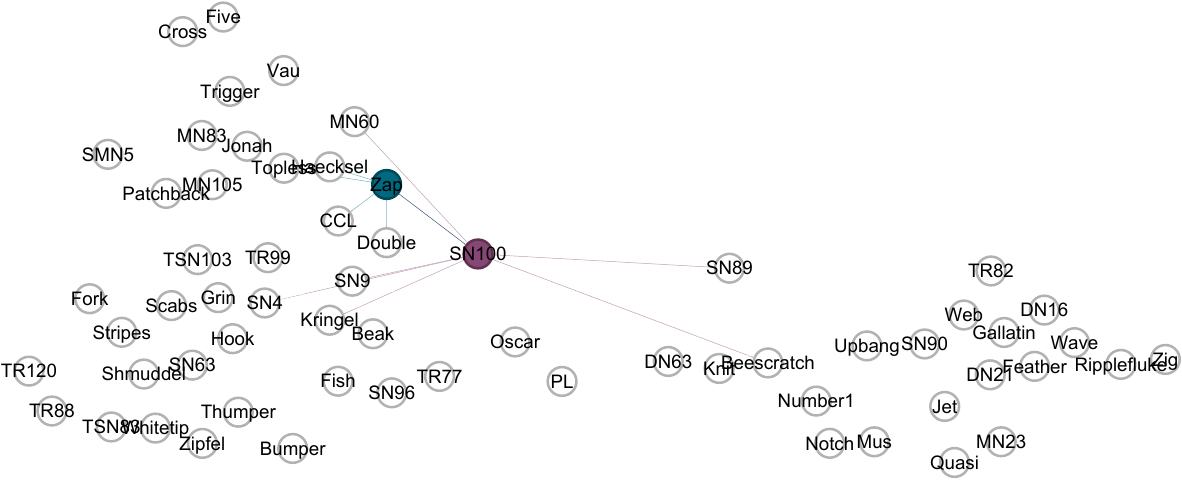}
    \vspace{2 mm}
    \label{fig:Do50_lo}
    \end{subfigure}
    \caption{The dolphin network overlapping nodes identified for three increasing threshold parameter values ($thr=0.0,1.0$, and $5.0$). The dolphin names are shown in the figure.}
    \label{fig:dolf}
\end{figure}

Table \ref{tab:dolphin_overlap_transposed} summarises our findings alongside a small selection of results from the literature. First, we examine the middle and bottom panels in Figure~\ref{fig:dolf}. The dolphins Beak, Kringel, MN105, Oscar, PL, and TR99 are consistent with the results reported in~\cite{Shen2009Overlap}. In contrast, SN4 and SN9 are not identified as overlapping nodes in our model, although they appear as overlapping nodes at a low threshold value in the upper panel of Figure~\ref{fig:dolf}. Our method identifies two groups: one consisting of five dolphins -- Beak, Bumber, Fish, SN96, and TR77, and another containing two -- Oscar and PL. The grouping in this way is similar to the findings in~\cite{Gopalan2017MixedMembership}. In our model, these groups remain together even when they change subcommunities, which is indicated by the brown shades in the figure. Furthermore, SN89 is identified in both cases, although it was not recognised in~\cite{Shen2009Overlap}. SN100 is still identified with a high threshold value in the bottom panel of Figure~\ref{fig:dolf}, highlighting its bridging role in the network structure. Zap is identified as an overlapping node across all threshold values, but not in the two models discussed in the literature.

\begin{table}[H]
\centering
\caption{Comparison of overlapping and boundary dolphins identified across selected studies.}
\begin{tabular}{l|c|c|c|c|c|}
& \multicolumn{5}{l}{\textbf{Study}} \\
\hline
 & \textbf{Lusseau \&} & \textbf{Shen} & \textbf{Gopalan} & \textbf{Our} & \textbf{Our} \\
 & \textbf{Newman} & et al. & et al. & \textbf{approach} & \textbf{approach} \\
\textbf{Dolphin} & (2004) \cite{Lusseau} & (2009) \cite{Shen2009Overlap} & (2017) \cite{Gopalan2017MixedMembership} & $thr = 1.0$ & $thr = 5.0$ \\
\hline
Beak &  & $\times$ & $\times$ & $\times$ &  \\
Bumper &  &  & $\times$ & $\times$ &  \\
Fish &  &  & $\times$ & $\times$ &  \\
Kringel &  & $\times$ &  & $\times$ &  \\
MN105 &  & $\times$ &  & $\times$ &  \\
Oscar &  & $\times$ & $\times$ & $\times$ &  \\
PL &  & $\times$ & $\times$ & $\times$ &  \\
SN4 &  & $\times$ &  &  &  \\
SN9 &  & $\times$ &  &  &  \\
SN89 &  &  & $\times$ & $\times$ &  \\
SN96 &  &  & $\times$ & $\times$ &  \\
SN100 & $\times$ &  &  & $\times$ & $\times$ \\
TR77 &  &  & $\times$ & $\times$ &  \\
TR99 &  & $\times$ &  & $\times$ &  \\
Zap &  &  &  & $\times$ & $\times$ \\

\end{tabular}
\label{tab:dolphin_overlap_transposed}
\end{table}

\subsection{Facebook Social Circles Network}
\label{subsec:Facebook}

The Facebook ego network of 4039 nodes and 88,234 edges is an undirected social network extracted from user friendship circles, first introduced in
~\cite{mcauley2012circles} as part of the SNAP dataset collection~\cite{data_SNAP}. Each node represents a Facebook user, and edges denote mutual friendships among friends of a single ego-centric user. The dataset includes ground-truth social circles that are derived through manual annotation, in which the ego user assigned friends to overlapping social groups based on real-life relationships and shared attributes, including school, workplace, location, or political affiliation. This makes it a popular benchmark for overlapping circle, community detection, and link prediction algorithms, including, for example, the methods in \cite{Yang2013BigCLAM} and \cite{Yang2012}. The analysis of these circles also raises an issue: The distinction between circles and communities is somewhat vague, and the terms are often used interchangeably in the science of social networks \cite{circles_communities2014}. This is problematic because circles, although they reflect some structural properties, e.g., triadic closures and cliques, are attribute driven, while communities are more a topological phenomenon \cite{Koistinen2025OverlappingNodes}. The overlapping community detection algorithms are often benchmarked against these circles and not the communities. Subsequent studies, including  \cite{shin2014circle-community}, have used this dataset to assess the quality of detected communities versus the memberships of known circles. The correlation between circles and communities is addressed in \cite{circles_communities2014}. Investigating how our detected overlapping nodes align with known circle intersections is left for future work.

\begin{figure}[h!]
    \centering
    \begin{subfigure}{0.45\textwidth}
        \includegraphics[width=\textwidth]{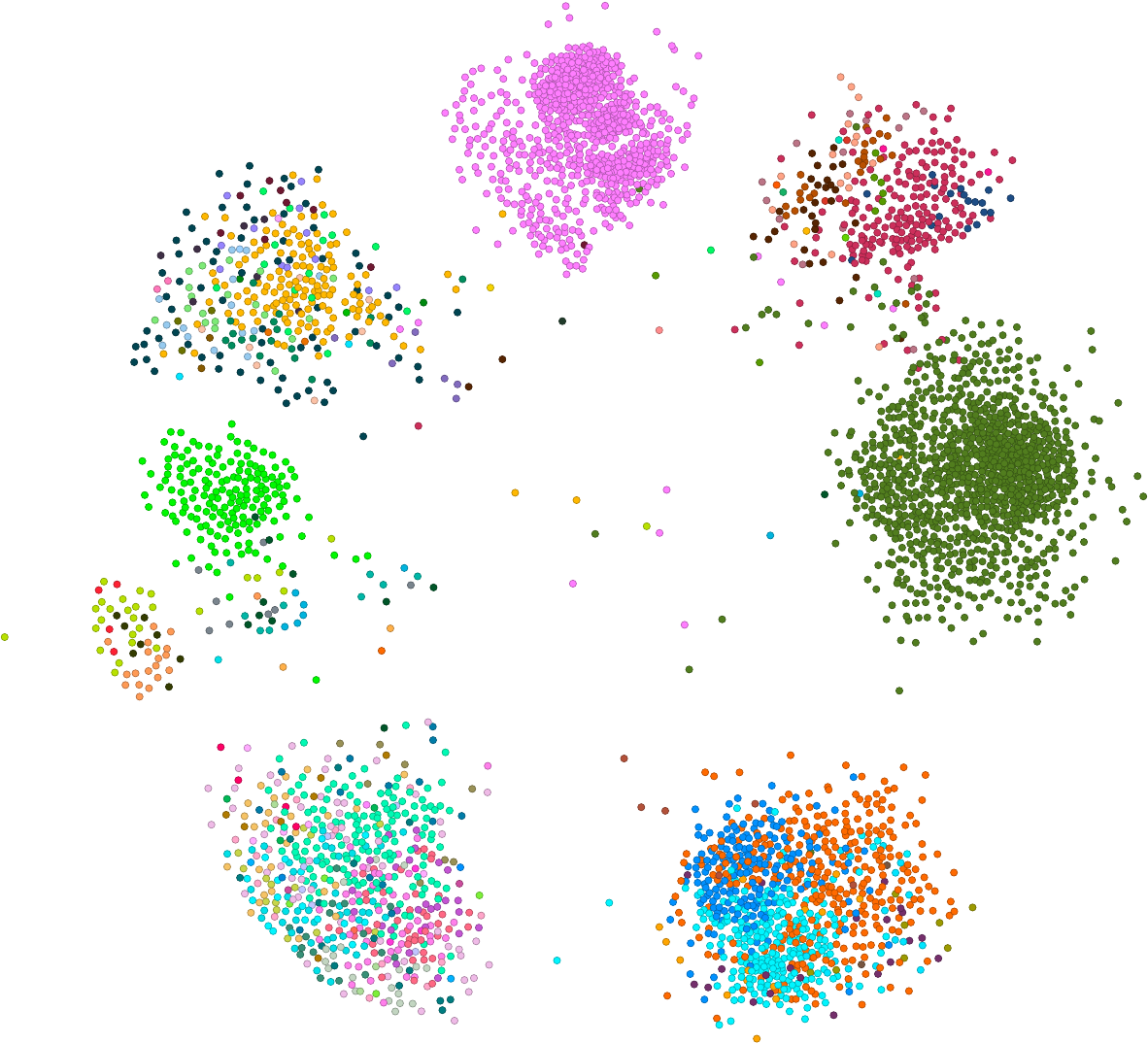}
        \label{fig:Fa00_lo_NE}
    \end{subfigure}
    \hfill
    \begin{subfigure}{0.45\textwidth}
        \includegraphics[width=\textwidth]{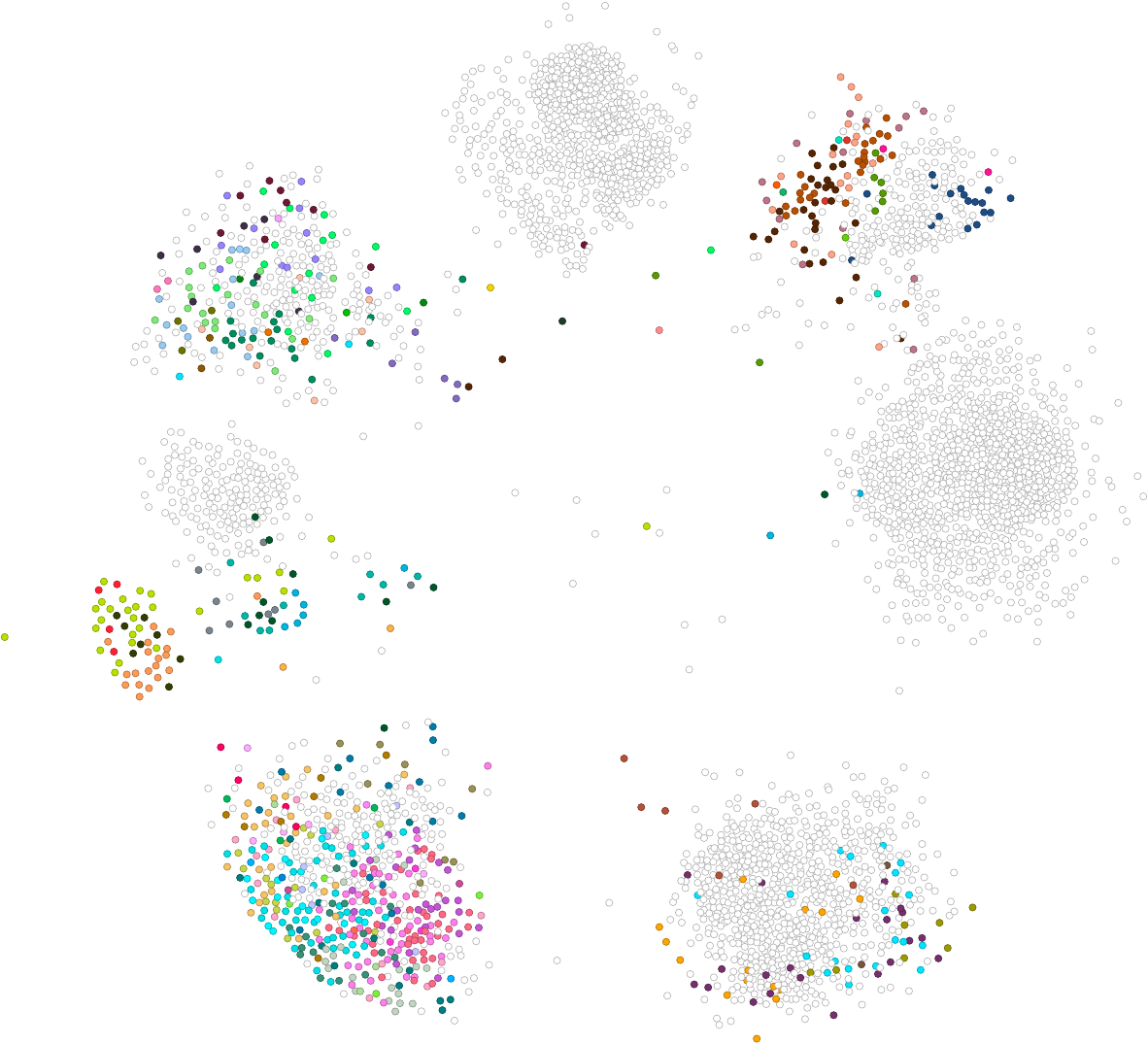}
        \label{fig:Fa01_lo_NE}
    \end{subfigure}

    \vspace{4mm}

    \begin{subfigure}{0.45\textwidth}
        \includegraphics[width=\textwidth]{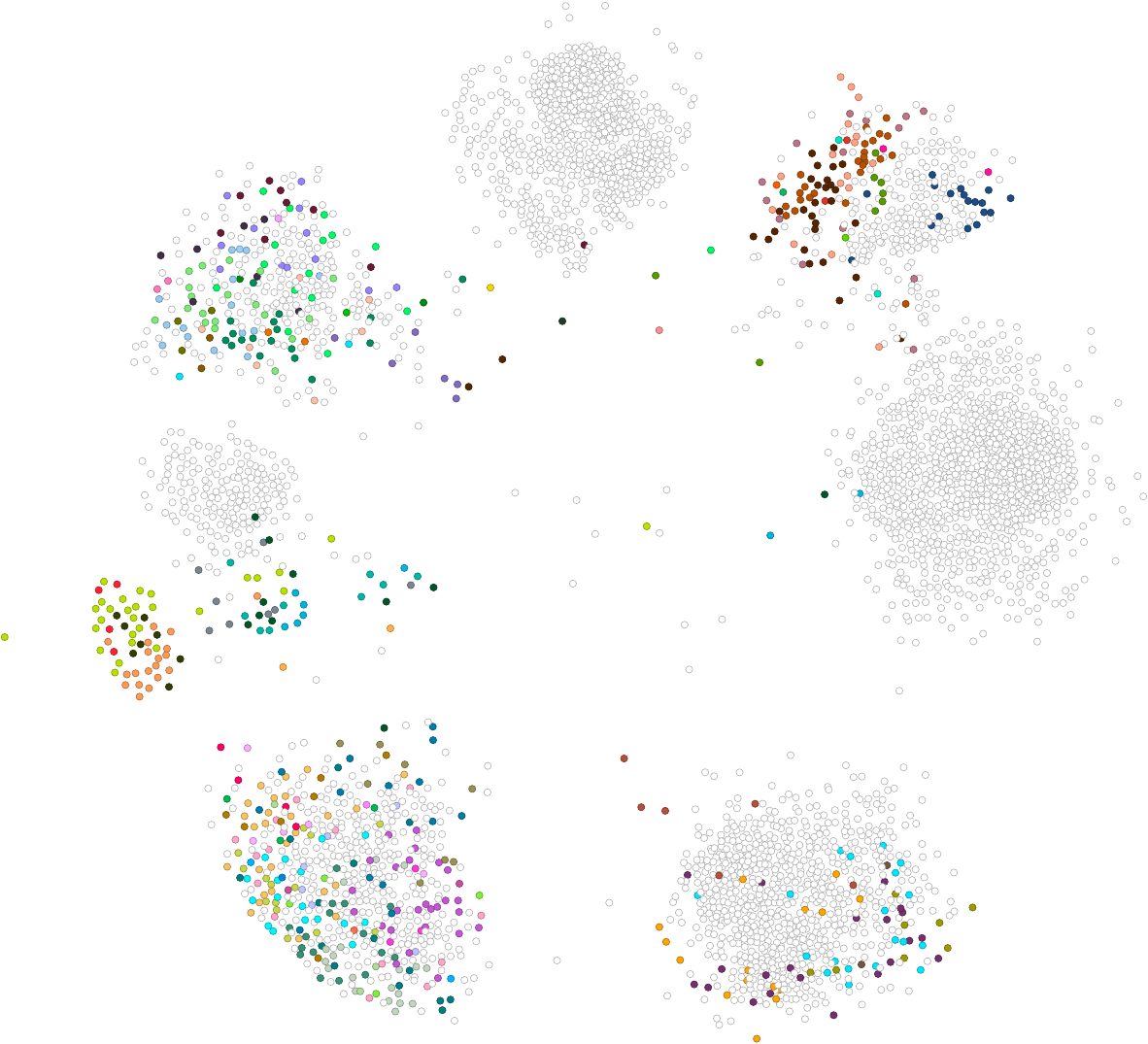}
        \label{fig:Fa02_lo_NE}
    \end{subfigure}
    \hfill
    \begin{subfigure}{0.45\textwidth}
        \includegraphics[width=\textwidth]{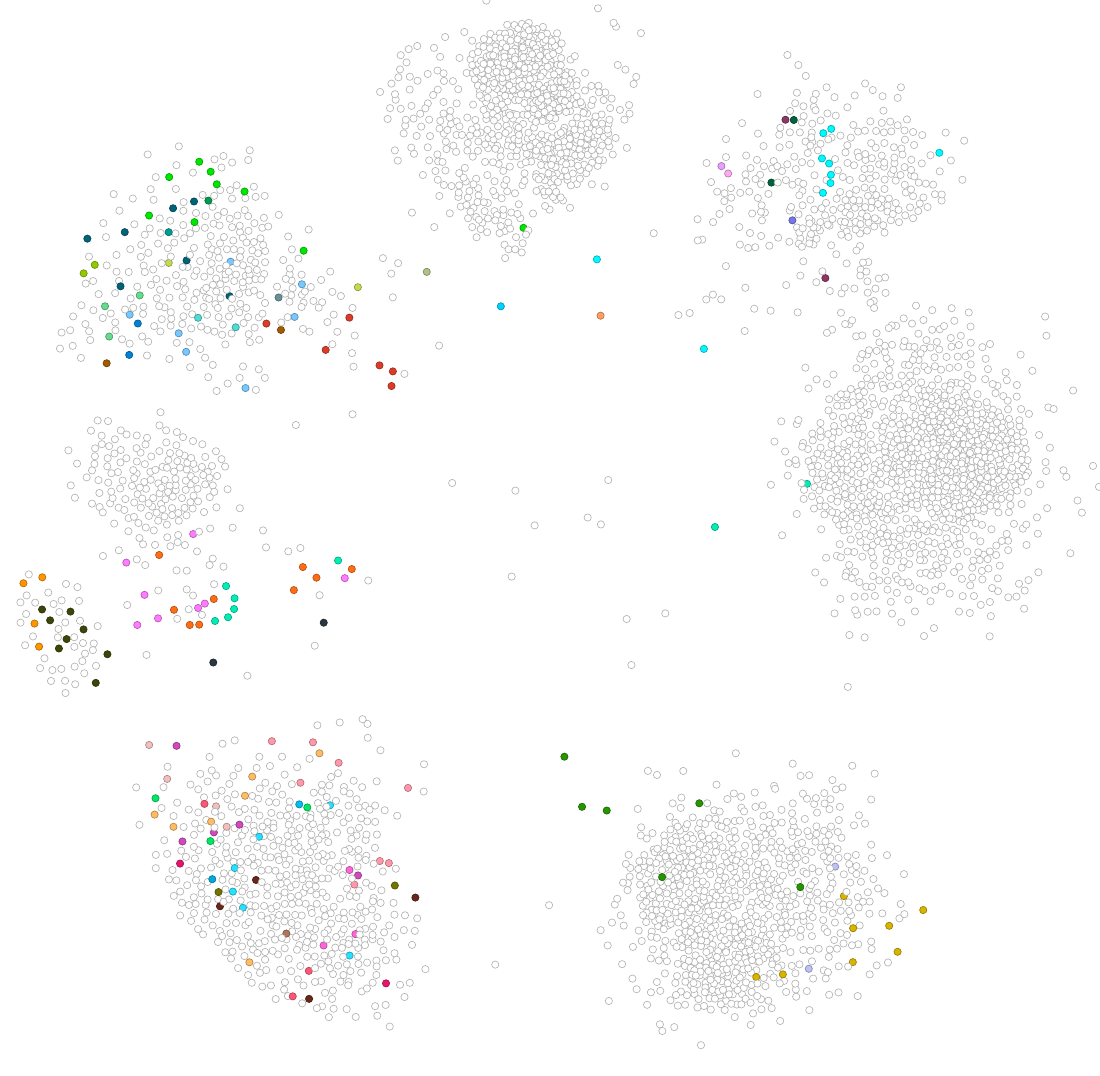}
        \label{fig:FB10_170e}
    \end{subfigure}

    \vspace{4mm}

    \begin{subfigure}{0.45\textwidth}
        \includegraphics[width=\textwidth]{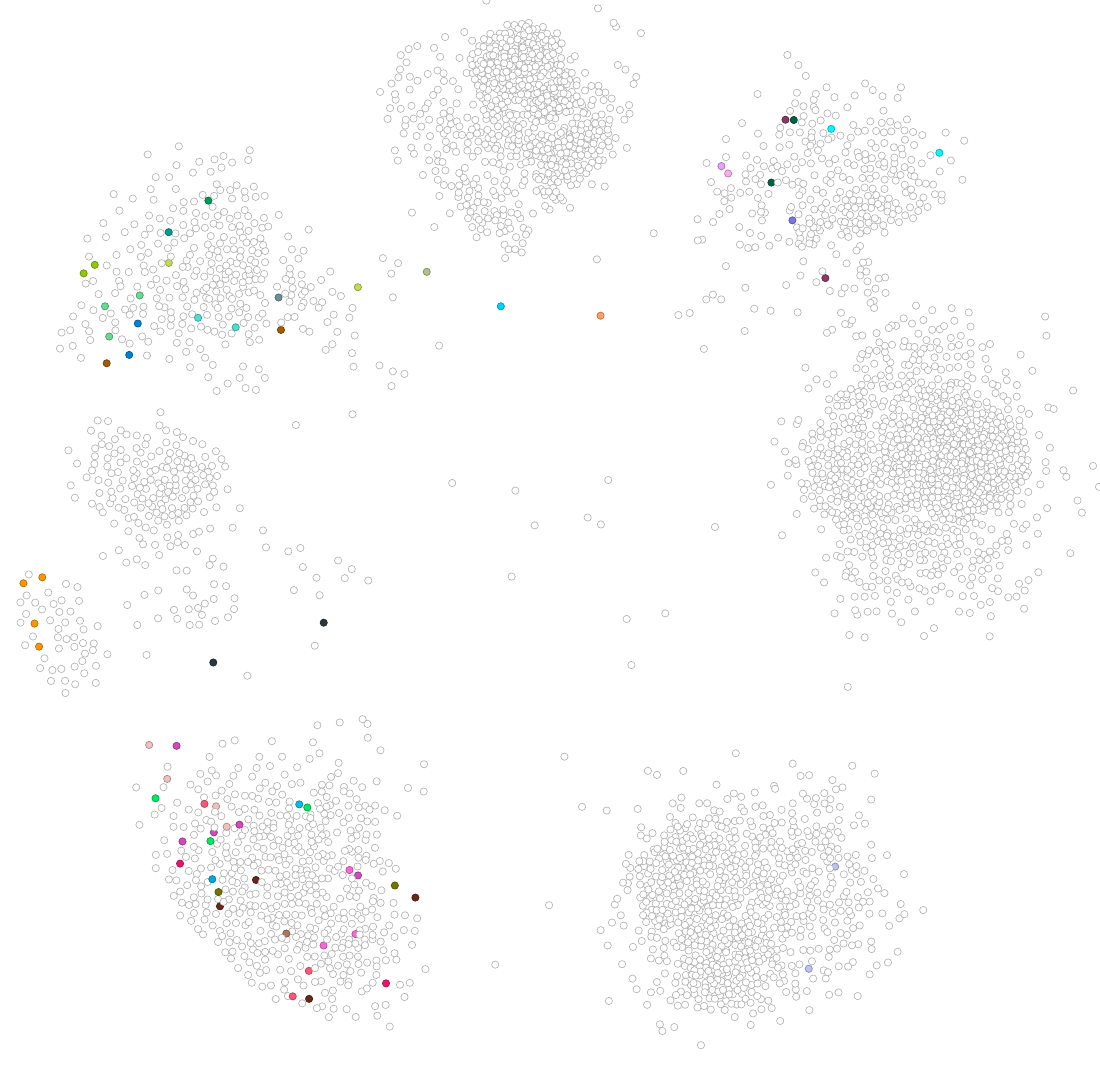}
        \label{fig:FB20_065e}
    \end{subfigure}
    \hfill
    \begin{subfigure}{0.45\textwidth}
        \includegraphics[width=\textwidth]{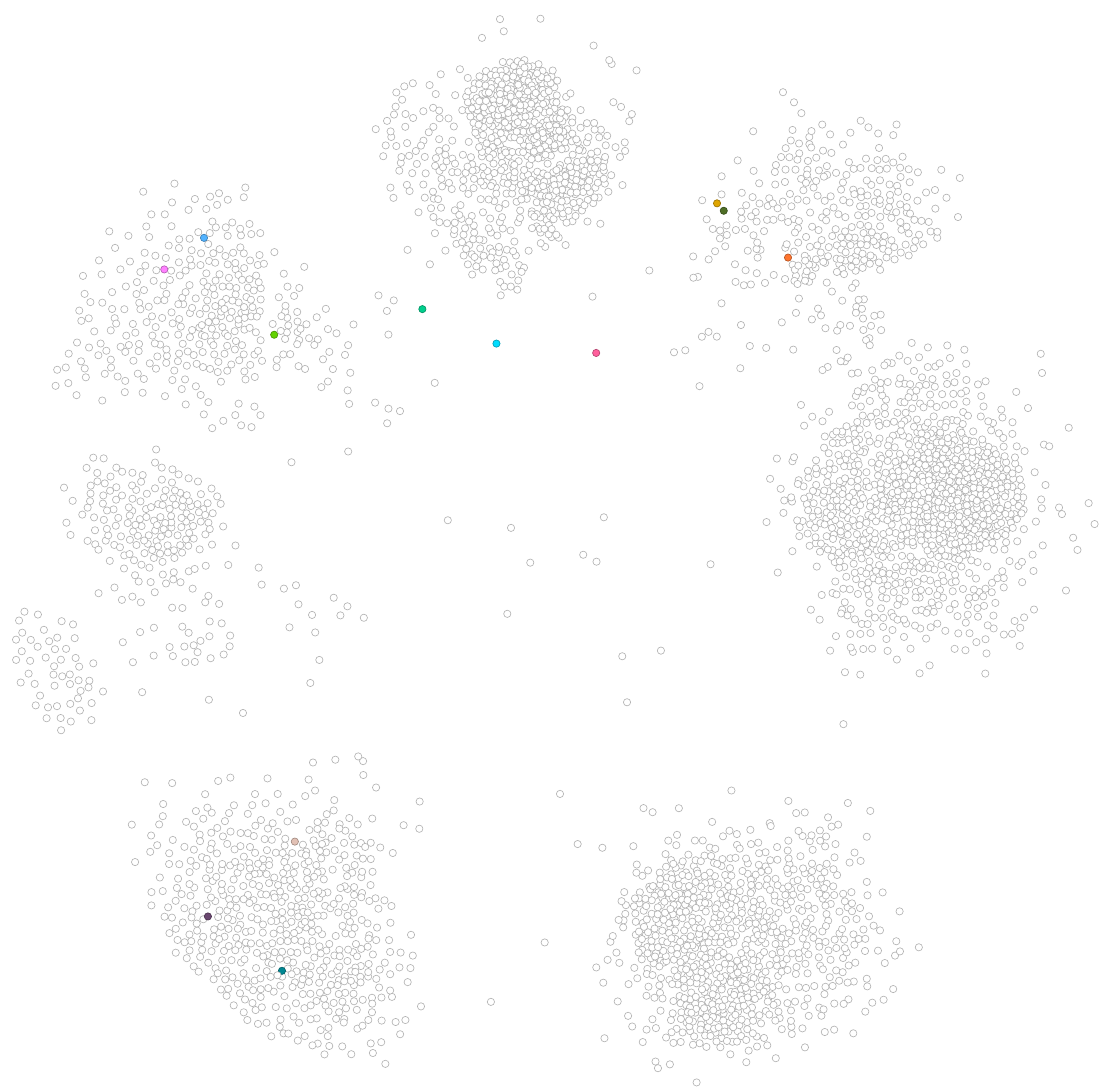}
        \label{fig:FB50_036e}
    \end{subfigure}

    \caption{Overlapping nodes detected in the Facebook network~\cite{data_SNAP}.
    The visualization shows the overlapping nodes identified at six different threshold parameter values.
    The rows correspond to thresholds $thr = (0, 1)$, $(2, 10)$, and $(20, 60)$.}
    \label{fig:Facebook}
\end{figure}

The Facebook ego-centric network serves as an example of a larger social network structure, illustrating our method for identifying overlapping nodes and groups of nodes. In Figure~\ref{fig:Facebook}, we show nodes that overlap in six different scenarios, each of which is represented by an increasing value of the threshold parameter. When the threshold is low, almost all nodes are classified as overlapping. This is because in dense networks, most nodes participate in multiple intersections, leading to a high overlapping rate. Increasing the threshold value reduces the number of overlapping nodes, ultimately resulting in only a few nodes being classified as overlapping within the network structure. Interestingly, as the threshold is increased, the nodes that remain classified as overlapping are not necessarily the ego nodes themselves; rather, they may be located anywhere within the network. This indicates that influential bridging nodes often reside within communities rather than at their boundaries, a pattern observed also, e.g., in \cite{Cha2010MeasuringUI}, in analysis of Twitter networks. Most importantly, our method can reveal such key bridging nodes even in densely interconnected subgroups. Such nodes would be difficult to identify using traditional community detection approaches.

\section{Discussion}
\label{sec:Discussion}

In this study, we focus on identifying and filtering overlapping nodes of detected overlapping communities. The key question is how to distinguish genuinely overlapping nodes from those that appear to overlap due to noise or highly similar community structures. Fine-grained methods for detecting overlapping communities often produce many closely related communities, especially in social networks where individuals belong to several social circles \cite{Shang2015}. In such dense settings, traditional methods may struggle to draw meaningful boundaries, and communities may share most of their members. In these cases, identifying true bridging nodes becomes more informative than analysing community structure alone.

Misclassifying too many nodes as overlapping obscures the influence of those that genuinely connect distinct network regions. Our approach addresses this by controlling structural noise with a threshold, reducing overlap caused by near-identical community compositions. This highlights nodes that function as real connectors and supports a more accurate analysis of influence, information flow, and key actors in complex networks. Furthermore, a topic not covered so far is the relation between the threshold parameter values and the network structure. Our empirical results indicate that the maximum threshold required to remove overlapping nodes is not arbitrary, but appears to be systematically related to the network characteristics including the size, density, and extent of local clustering (see Table \ref{tab:Networks}. Small and moderately sized social networks such as Les Misérables, the Zachary's Karate Club, and Dolphin Social Networks require only low to moderate thresholds, reflecting their relatively simple community structures and limited community overlap. In contrast, the large and dense Facebook network requires a much higher threshold, implying a far richer pattern of overlapping communities. Subsequently, the similarly large but very sparse Mobile Phone Network exhibits the lowest threshold, suggesting that sparsity suppresses the overlap despite scale. Although these findings are preliminary and subject to future work, they empirically suggest that the interplay between network size, density, and clustering jointly governs the sensitivity of overlapping nodes to the chosen threshold parameter.

Ego-centric networks are typically analysed by focusing on a single ego and its alters, which can challenge the capability of community detection algorithms. In our analysis of the Facebook ego-centric network, we have used a relatively low link weight to identify all major communities as distinct (Figure~\ref{fig:Facebook}). This produced numerous communities with only minor overlaps, making bridging nodes harder to detect. The number of overlapping nodes can be controlled by fine-tuning the threshold: as it increases, the number of overlapping nodes decreases. This shift from a granular to a clearer structural representation aids the interpretation of the community organisation and influences the dynamics, as also demonstrated in the mobile phone call network.

In future algorithm development, we can introduce selection rules that are specifically tailored to meet the needs of different applications. These rules could take into account factors such as the size and coherence of intersections between communities, the number of communities to which a node belongs, and the strength of those memberships. By incorporating these criteria, the algorithm could better prioritise overlapping behaviours, allowing it to identify key bridging nodes and model patterns of multi-membership. A further point concerns the threshold rule applied in our approach. Although it serves as a practical heuristic, it lacks a solid theoretical basis. This limitation reflects a broader issue in community detection: there is no universally accepted definition of what constitutes a community. The absence of a clear conceptual foundation complicates the development of a cohesive theoretical framework for any method that detects overlapping nodes. Therefore, we argue that selection criteria should be based on specific use cases, since different analytical goals lead to various ways of prioritising overlaps.

\section{Conclusions}
\label{sec:Conclusions}

This study introduced a framework for classifying shared nodes within overlapping communities of complex networks. By differentiating overlapping nodes from noise or subcommunity structures, our method improves the interpretation of community overlaps and reduces the ambiguity arising from nearly identical community divisions. The approach integrates a selection rule and a tuneable threshold parameter to adjust the level of overlap, allowing the analysts to concentrate on structually the most significant bridging nodes.

Applications to several real-world networks, based on social, communication, and empirical datasets, demonstrated that increasing the threshold progressively filters out spurious overlaps while preserving core connector nodes. These results confirm the robustness and adaptability of our method across various types of networks. Importantly, the approach is compatible with any community detection algorithm capable of producing overlapping partitions, thus providing a flexible post-processing tool for network analysis.

Future studies should focus on investigating how sensitive the results are to different threshold choices, as their effects are influenced by community structures generated by specific detection methods. Variations in threshold levels can also serve as a diagnostic tool to better understand both the behaviour of the detection methods and the underlying network structure. Since the framework is not limited to any particular algorithm—our own model served only as an example—it would be beneficial to evaluate it using additional overlapping community detection techniques. Conducting benchmarks on synthetic or ground-truth networks, along with analyses that link identified overlapping nodes to such metrics as influence or centrality, will deepen our understanding of when the framework provides the most insightful perspectives on overlapping structures.

\pagebreak[4]

\appendix
\section[\appendixname~\thesection]{}
\label{sec:AppendiA}

\subsection*{Building Blocks}
\label{Appendix:BB}

These algorithms evaluate the input community structures by comparing them with a set of previously detected network divisions, each consisting of two communities. It first computes a cohesion score for each division, representing the strength of influence or connectivity among the nodes within each community (Algorithm \ref{alg:compute-rank-scores}). Then these divisions are sorted by their cohesion values. From the top divisions, the algorithm constructs building blocks that are used to classify nodes in the input communities (Algorithm \ref{alg:build-blocks}).

\begin{algorithm}[H]
\caption{ComputeAndRankPartition}
\label{alg:compute-rank-scores}
\begin{algorithmic}[1]
\Require
\Statex $N \in \mathbb{N}$ \Comment{number of nodes}
\Statex $M \in \mathbb{R}^{N \times N}$ \Comment{weight matrix}
\Statex $T \in \mathbb{N},\ T \ge 2$ \Comment{number of iterations/partitions}
\Statex $Z \in \{0,1\}^{T \times N}$ \Comment{row $t$ is the indicator vector at iteration $t$}
\Statex $\texttt{normalize} \in \{\mathbf{true},\mathbf{false}\}$ \Comment{whether to apply normalization}
\Ensure
\Statex $s \in \mathbb{R}^{T-1}$ \Comment{agreement score per iteration}
\Statex $\pi$ \Comment{indices $\{1,\dots,T-1\}$ sorted by $s$ in descending order}

\Statex

\State \textbf{Initialization}
\State $s \gets \mathbf{0} \in \mathbb{R}^{T-1}$
\State $\pi \gets [1,2,\dots,T-1]$

\For{$t = 1$ to $T-1$}
    \State $v \gets Z[t+1,:] \in \{0,1\}^{N}$ \Comment{indicator at iteration $t{+}1$}
    \State $\texttt{total} \gets 0$
    \For{$i = 1$ to $N$}
        \State \(\displaystyle \text{contrib}(i) \gets \sum_{j=1}^{N}(M_{ij}+M_{ji})\big[v_j v_i + (1-v_j)(1-v_i)\big]\)
        \State $\texttt{total} \gets \texttt{total} + \text{contrib}(i)$
    \EndFor
    \If{$\texttt{normalize}$}
        \State \(\displaystyle p_0 \gets \frac{N - \sum_{j=1}^N v_j}{N}\) \Comment{fraction of zeros in $v$}
        \State \(\displaystyle \texttt{total} \gets \frac{\texttt{total}}{1 - 2\,p_0(1-p_0)}\)
    \EndIf
    \State \(s[t] \gets \texttt{total}/2\) \Comment{avoid symmetric double-counting}
\EndFor

\State \textbf{Sort} $\pi$ by $s$ in descending order
\end{algorithmic}
\end{algorithm}

\begin{algorithm}[H]
\caption{BuildBlocks}
\label{alg:build-blocks}
\begin{algorithmic}[1]
\Require
\Statex $N \in \mathbb{N}$ \Comment{number of nodes}
\Statex $Z \in \{0,1\}^{T \times N}$ \Comment{membership indicators over $T$ iterations}
\Statex $T \in \mathbb{N},\ T \ge 2$ \Comment{number of iterations}
\Statex $\texttt{segment\_length} \in \mathbb{N}$ \Comment{number of iterations in each segment}
\Statex $\pi$ \Comment{ranking from Algorithm~\ref{alg:compute-rank-scores} over $t=1,\dots,T-1$}
\Statex $K \in \mathbb{N}$ \Comment{number of top-ranked segment anchors to process}
\Ensure
\Statex For each anchor $r \in \{1,\dots,K\}$ and segment index $\texttt{seg}$:
\Statex \hspace{1.5em} $\mathcal{B}^{(r,\texttt{seg})}$: distinct membership-signature blocks
\Statex \hspace{1.5em} $a^{(r,\texttt{seg})}$: node $\rightarrow$ block assignments
\Statex \hspace{1.5em} $\texttt{block\_sizes}^{(r,\texttt{seg})}$: block sizes

\Statex


\For{$r = 1$ to $K$} \Comment{process top-$K$ ranked anchors}
    \State $t^\star \gets \pi[r]$ \Comment{score index from Algorithm 1}
    \State $\texttt{segment\_end} \gets t^\star + 1$
    \For{$\texttt{seg} = \texttt{segment\_end} - \texttt{segment\_length} + 1$ to $\texttt{segment\_end} + 1$}
        \State $t_1 \gets T - \texttt{seg} + 3$; \quad $t_2 \gets T$
        \State $\mathcal{B} \gets \emptyset$; \quad $\texttt{num\_blocks} \gets 0$
        \State $a \gets \mathbf{0} \in \mathbb{N}^{N}$
        \State $\texttt{block\_sizes} \gets \mathbf{0} \in \mathbb{N}^{N}$

        \For{$j = 1$ to $N$}
            \State $\texttt{sig} \gets (Z[t,j])_{t=t_1}^{t_2}\ $ \Comment{Set the Pattern} 
            \State $\texttt{b\_idx} \gets$ index of a block in $\mathcal{B}$ equal to $\texttt{sig}$ (if any; else $0$)
            
            \If{$\texttt{b\_idx} \neq 0$}
                \State $\texttt{block\_sizes}[\texttt{b\_idx}] \gets \texttt{block\_sizes}[\texttt{b\_idx}] + 1$; \quad $a[j] \gets \texttt{b\_idx}$
            \Else
                \State $\texttt{num\_blocks} \gets \texttt{num\_blocks} + 1$
                \State Add $B_{\texttt{num\_blocks}} \gets \texttt{sig}$ to $\mathcal{B}$
                \State $\texttt{block\_sizes}[\texttt{num\_blocks}] \gets 1$; \quad $a[j] \gets \texttt{num\_blocks}$
            \EndIf
        \EndFor

        \State Output $(\mathcal{B}, a, \texttt{block\_sizes})$ for anchor $r$ and segment $\texttt{seg}$
    \EndFor
\EndFor
\end{algorithmic}
\end{algorithm}

\newpage

\subsection*{Patterns}
\label{Appendix:BB2}

The table below illustrates how the $34$ nodes, \( N = 1, \ldots, 34 \), of the Karate Club network are categorised into patterns (used in Algorithm~\ref{alg:build-blocks}).

{\scriptsize
\begin{longtable}{c|c|c|c|c|c|c|c|c}
\toprule
N & Div1 & Div2 & Div3 & Div4 & Div5 & Div6 & Div7 & Pattern \\
\midrule
1  & x & x & x & x & x & x & x & xxxxxxx \\
2  & x & x & x & x & x & x & x & xxxxxxx \\
3  & x & x & x & x & o & x & x & xxxxoxx \\
4  & x & x & x & x & x & x & x & xxxxxxx \\
5  & o & x & x & x & o & o & o & oxxxooo \\
6  & o & x & x & x & o & o & o & oxxxooo \\
7  & o & x & x & x & o & o & o & oxxxooo \\
8  & x & x & x & x & x & x & x & xxxxxxx \\
9  & x & o & o & x & o & o & x & xooxoox \\
10 & x & o & o & x & o & o & x & xooxoox \\
11 & o & x & x & x & o & o & o & oxxxooo \\
12 & x & x & x & x & x & x & x & xxxxxxx \\
13 & x & x & x & x & x & x & x & xxxxxxx \\
14 & x & x & x & x & x & x & x & xxxxxxx \\
15 & x & o & o & o & o & o & o & xoooooo \\
16 & x & o & o & o & o & o & o & xoooooo \\
17 & o & x & x & x & o & o & o & oxxxooo \\
18 & x & x & x & x & x & x & x & xxxxxxx \\
19 & x & o & o & o & o & o & o & xoooooo \\
20 & x & x & x & x & x & x & x & xxxxxxx \\
21 & x & o & o & o & o & o & o & xoooooo \\
22 & x & x & x & x & x & x & x & xxxxxxx \\
23 & x & o & o & o & o & o & o & xoooooo \\
24 & x & o & o & o & o & o & o & xoooooo \\
25 & x & o & x & x & o & x & x & xoxxoxx \\
26 & x & o & x & x & o & x & x & xoxxoxx \\
27 & x & o & o & o & o & o & o & xoooooo \\
28 & x & o & o & x & o & o & x & xooxoox \\
29 & x & o & x & x & o & x & x & xoxxoxx \\
30 & x & o & o & o & o & o & o & xoooooo \\
31 & x & o & o & x & o & o & x & xooxoox \\
32 & x & o & x & x & o & x & x & xoxxoxx \\
33 & x & o & o & o & o & o & o & xoooooo \\
34 & x & o & o & o & o & o & o & xoooooo \\
\bottomrule
\end{longtable}}

Seven network divisions, denoted \(Div\), have been identified using a community detection method. Partitions are indicated by ‘x’ and ‘o’ for each node, where ‘x’ denotes membership in the partition and ‘o’ indicates non-membership (or membership in the two distinct communities). The cumulative patterns derived from these columns are displayed in the last column of the table. In the first pattern, 'xxxxxxx', there are ten nodes: {1, 2, 4, 8, 12, 13, 14, 18, 20, 22}. The second pattern, 'xxxxoxx', contains one node: {3}. Additional patterns continue in this manner. In particular, the nodes identified in patterns 2, 4, and 6 overlap when the threshold \(thr = 0\). Additionally, node {3} overlaps at a threshold of \(thr = 0.5\) (according to Algorithm~\ref{alg:community-eval-using-blocks}).

{\small
\begin{table}[H]
    \centering
    \begin{tabular}{|c|c|c|}\hline
 Pattern no& Pattern&No nodes\\\hline\hline
        1 & xxxxxxx & 10 \\\hline
        2 & xxxxoxx & 1 \\\hline
        3 & oxxxooo & 5 \\\hline
        4 & xooxoox & 4 \\\hline
        5 & xoooooo & 10 \\\hline
        6 & xoxxoxx & 4 \\ \hline
    \end{tabular}
    \caption{The six patterns found in the Karate Club network.}
    \label{tab:placeholder}
\end{table}}

\pagebreak[4]

\section[\appendixname~\thesection]{}
\label{Appendix:OverlappingNodes}
\subsection*{Overlapping Nodes}

Each input community is analysed by assigning its nodes to building blocks and checking their alignment with the detected patterns using a tolerance threshold. Unmatched nodes are identified, i.e., overlapping nodes are identified.

\setcounter{algorithm}{0}
\renewcommand{\thealgorithm}{B\arabic{algorithm}}
\begin{algorithm}[H]
\caption{CommunityEvaluation}
\label{alg:community-eval-using-blocks}
\begin{algorithmic}[1]
\Require
\Statex $N \in \mathbb{N}$ \Comment{number of nodes}
\Statex $K \in \mathbb{N}$ \Comment{number of communities}
\Statex $c[1\ldots N] \in \{1,\dots,K\}$ \Comment{community assignment per node}
\Statex $r \in \mathbb{N}$, $\texttt{seg} \in \mathbb{N}$ \Comment{anchor and segment index chosen from Algorithm~\ref{alg:build-blocks}}
\Statex $a^{(r,\texttt{seg})}[1\ldots N] \in \mathbb{N}$ \Comment{node $\rightarrow$ block assignment for the chosen $(r,\texttt{seg})$}
\Statex $\texttt{block\_sizes}^{(r,\texttt{seg})}[1\ldots B] \in \mathbb{N}$ \Comment{sizes of the $B$ distinct blocks for $(r,\texttt{seg})$}
\Statex $thr \in \mathbb{R}_{\ge 0}$ \Comment{Threshold-parameter}
\Statex $\gamma[1\ldots K]$ \Comment{output flags: match (=1) or not matched (=-1)}
\Statex $o[1\ldots N]$ \Comment{accumulator for nodes in communities that fail matching}
\Ensure
\Statex $\mathcal{O} = \{\, i : o[i] = 1 \,\}$ \Comment{Set of nodes belonging to communities not matching any block}
\Statex $|\mathcal{O}|$ \Comment{Number of such nodes (total size of unmatched communities)}

\Statex

\State $B \gets \operatorname{len}(\texttt{block\_sizes}^{(r,\texttt{seg})})$
\State Compute community sizes:
\For{$b = 1$ to $K$}
    \State $\texttt{comm\_size}[b] \gets \left|\{\, i : c[i]=b \,\}\right|$
\EndFor
\State $\gamma[1\ldots B] \gets -1$ \Comment{initialize all blocks as unmatched}
\State $o[1\ldots N] \gets 0$ \Comment{initialize accumulator}

\For{$k = 1$ \textbf{to} $B$}
    \State $v[1\ldots N] \gets 0$ \Comment{indicator for nodes in block $k$}
    \For{$i = 1$ \textbf{to} $N$}
        \If{$a^{(r,\texttt{seg})}[i] = k$} 
            \State $v[i] \gets 1$
        \EndIf
    \EndFor
    \State $S \gets \sum_{j=1}^{N} v[j]$ \Comment{size of block $k$}

    \For{$b = 1$ \textbf{to} $K$} \Comment{iterate over communities from Algorithm~\ref{alg:build-blocks}}
        \State $T_1 \gets \texttt{comm\_size}[b]$ \Comment{community size}

       \If{$S \neq 0$}
    \If{$\frac{|T_1 - S|}{S} \ge thr \;\textbf{or}\; \frac{|(N - T_1) - S|}{S} \ge thr$}
        \State $\gamma[k] \gets -1$ \Comment{no match on this block}
    \Else
            \State $\gamma[k] \gets 1$ \Comment{community matches this block}
        \State \textbf{break}

    \EndIf
\Else
    \State $\gamma[k] \gets 1$ \Comment{(S = 0)}
\EndIf

        \If{$b = k$ \textbf{and} $\gamma[k] = -1$}
            \State $o \gets o + v$ \Comment{accumulate nodes of unmatched community}
        \EndIf
    \EndFor
\EndFor

\State $\mathcal{O} \gets \{\,i \in \{1,\dots,N\} : o[i] = 1\,\}$ \Comment{final set of overlapping nodes}
\State $|\mathcal{O}| \gets \sum o$

\end{algorithmic}
\end{algorithm}

\newpage

\newpage

\section*{Conflicts of Interest}
The authors declare no conflict of interest.
\section*{Author Contributions}
Conceptualization, V.K. and K.K.; Methodology, V.K. and K.K.; Software, V.K. and K.K.; Validation, V.K. and K.K.; Formal Analysis, V.K. and K.K.; Writing – Original Draft Preparation, V.K. and K.K.; Writing – Review \& Editing, V.K., K.K. and K.K.K.; Visualization, V.K. and K.K.; Supervision, K.K.K:.; Project Administration,K.K.K:; Funding Acquisition, K.K.K.

\section*{Data Availability}
The Zachary's Karate Club friendship network used in Section~\ref{subsec:Karate} is a standard benchmark dataset in community detection research. It is publicly available, for example, in
\href{https://networkrepository.com/soc-karate.php}{https://networkrepository.com/soc-karate.php}.
\\
\\
In Section~\ref{subsec:Kurjat}, we examine the \emph{Les Misérables} character co-appearance network. It is publicly available from repositories such as 
\href{https://networkrepository.com/lesmis.php}{https://networkrepository.com/lesmis.php} and appears in multiple open-source network libraries.
\\
\\
The mobile phone call network studied in Section~\ref{subsec:Mobile} is proprietary and cannot be shared publicly due to contractual and privacy restrictions. Access may be granted only under a data-sharing agreement with the provider and subject to applicable ethical and legal constraints.
\\
\\
The bottlenose dolphin social network analysed in Section~\ref{subsec:Dolphins} is publicly available from several online repositories, including the 
\href{https://networkrepository.com/soc-dolphins.php}{https://networkrepository.com/soc-dolphins.php}.
\\
\\
The ego-Facebook social circles network analysed in Section~\ref{subsec:Facebook} is publicly available through the Stanford Large Network Dataset Collection (SNAP):  
\href{https://snap.stanford.edu/data/ego-Facebook.html}{https://snap.stanford.edu/data/ego-Facebook.html}.

\pagebreak[3]

\bibliographystyle{plain}   

\bibliography{bib.bib}


\end{document}